\documentclass[a4paper,11pt]{article}
\pdfoutput=1

\usepackage{jcappub}

\usepackage{graphicx}
\usepackage{amsmath}
\usepackage{amssymb}
\usepackage{amsfonts}
\usepackage{dcolumn}
\usepackage{bm}
\usepackage[dvipsnames]{xcolor}
\usepackage[utf8]{inputenc}
\usepackage{subfigure}
\usepackage{gensymb}
\usepackage{cleveref}
\usepackage{colortbl}
\usepackage{tabularx}
\usepackage{tabulary}

\usepackage[T1]{fontenc}
\usepackage{pstricks}
\usepackage{color}
\usepackage{multirow}
\usepackage{slashed}
\usepackage{mathtools}

\usepackage[force]{feynmp-auto}
\usepackage{verbatim}

\newcommand{\Hub}{\mathcal{H}}

\hypersetup{colorlinks,linkcolor={blue},citecolor={teal},urlcolor={violet}}

\newcommand{\SOTON}{Department of Physics and Astronomy, University of Southampton, SO17 1BJ Southampton, United Kingdom}

\begin{document}

\title{Gravitational wave signals from leptoquark-induced first-order electroweak phase transitions}

\author[a,1]{B. Fu,\note{\url{https://orcid.org/0000-0003-2270-8352}}}
\author[a,2]{S. F. King\note{\url{https://orcid.org/0000-0002-4351-7507}}}

\affiliation[a]{\SOTON}

\emailAdd{B.Fu@soton.ac.uk}
\emailAdd{king@soton.ac.uk}

\date{\today}

\abstract{
We consider the extension of the Standard Model (SM) with scalar leptoquarks in $SU(2)$ singlet, doublet and triplet representations. Through the coupling between leptoquark and the SM Higgs field, the electroweak phase transition (EWPT) can turn into first-order and consequently produce gravitational wave signals. We compute the required value of the leptoquark-Higgs for first-order EWPT to happen and discuss about the possible constraint from Higgs phenomenology. Choosing some benchmarks, we present the strength of the gravitational waves produced during the leptoquark-induced first-order EWPT and compare them to detector sensitivities. We find that the $SU(2)$ representations of the leptoquark can be distinguished by gravitational waves in the parameter space where first-order EWPT can happen as a function of the Higgs portal coupling. 
}

\maketitle
\tableofcontents

%%%%%%%%%%%%%%%%%%%%%%%%%%%%%%%%%%%%%%%%%%%%%%%%%%%%%%%%%%%%%%%%%%%%%
\section{Introduction}
%%%%%%%%%%%%%%%%%%%%%%%%%%%%%%%%%%%%%%%%%%%%%%%%%%%%%%%%%%%%%%%%%%%%%
Leptoquarks (LQs) are hypothetical particles that can convert quarks into leptons and vice versa with great interest in elementary particle physics. From the theoretical aspect, it has been predicted naturally by the Pati-Salam unification of quarks and leptons \cite{Pati:1973uk,Pati:1974yy}, where the leptoquark is first raised, as well as many other grand unified theories \cite{Senjanovic:1982ex,Buchmuller:1986iq,Frampton:1991ay,Gershtein:1999gp,Fuentes-Martin:2020pww,King:2021jeo,FernandezNavarro:2022gst}. From the experimental side, the existence of leptoquarks is strongly indicated by lepton flavour universality violation (LFUV) in semi-leptonic $B$ decay \cite{LHCb:2019hip,Belle:2019rba,LHCb:2021trn,LHCb:2021lvy,Belle:2019rba,Angelescu:2021lln,Becirevic:2022tsj}. Besides LFUV, leptoquarks can also be related to a wide variety of phenomena beyond the standard model, including the muon $g-2$ \cite{Cheung:2001ip,ColuccioLeskow:2016dox,Crivellin:2019dwb,Athron:2021iuf,Du:2021zkq,Chen:2022hle}, the neutrino mass \cite{Mahanta:1999xd,Deppisch:2016qqd,Popov:2016fzr,Cai:2017wry,BhupalDev:2020zcy} and the $W$ boson mass \cite{Crivellin:2020ukd,DAlise:2022ypp,Athron:2022qpo,Cheung:2022zsb,Bhaskar:2022vgk,He:2022zjz}.

Despite the theoretical and experimental attraction from leptoquarks, they have not been found by any collider experiment so far. One of the possibilities to find leptoquark is through its connection to Higgs phenomenology \cite{Kolb:1997rb,Dorsner:2016wpm}. Generically, the scalar leptoquarks can couple to Higgs boson in the scalar potential. After electroweak symmetry breaking, the leptoquark-Higgs operator induces the couplings to the physical Higgs boson, which can further affect loop-induced Higgs production and decay processes. Such effects can be probed with the Higgs signal strength measurements at colliders and thus are potential smoking guns for leptoquarks.  

In the meantime, the Higgs portal allows leptoquarks to modify the EWPT in the early universe. It has been shown that first-order EWPT can be induced by an additional singlet scalar field without any vacuum expectation value (VEV) \cite{Curtin:2014jma}. And the stochastic gravitational wave background produced during the cosmological phase transition can be potentially tested by detectors \cite{Caprini:2015zlo,Weir:2017wfa,Caprini:2019egz}. This provides us with a new possibility of testing scalar leptoquarks, using a similar approach to the singlet, from cosmic signals.

In this paper, we extend the study of first-order EWPT induced by an extra singlet scalar to the case of scalar leptoquarks in $SU(2)$ singlet, doublet and triplet representations, and show how such leptoquarks can affect the EWPT through their coupling to the standard model Higgs boson. By computing the effective scalar potential, we find the range of Higgs portal coupling where eligible first-order EWPT can happen for different types of scalar leptoquark with a mass around TeV scale. Then we apply the standard procedure \cite{Beniwal:2017eik,DiBari:2021dri,Bian:2021dmp,Demidov:2021lyo,Graf:2021xku,Zhou:2022mlz,Chen:2022zsh,Lu:2022zpn,Dasgupta:2022isg} to calculate the gravitational wave background produced during the first-order EWPT induced by leptoquark for some benchmark cases and compare it with the detector sensitivities. We found that in some range of the parameter space, the leptoquark-induced first-order EWPT is able to produce gravitational wave signals that are strong enough to be detected. 

The paper is organised as follows. In Sec.\ref{sec:1stPT}, we discuss the first-order EWPT induced by leptoquark through the Higgs portal. We also show the constraints from Higgs physics to the parameter space. In Sec.\ref{sec:GW}, we show the gravitation wave signal produced during leptoquark-induced first-order EWPT for benchmark points. Finally, we summarise and conclude in Sec.\ref{sec:con}.

%%%%%%%%%%%%%%%%%%%%%%%%%%%%%%%%%%%%%%%%%%%%%%%%%%%%%%%%%%%%%%%%%%%%%
\section{first-order EWPT induced by scalar leptoquarks \label{sec:1stPT}}
%%%%%%%%%%%%%%%%%%%%%%%%%%%%%%%%%%%%%%%%%%%%%%%%%%%%%%%%%%%%%%%%%%%%%
In this section, we discuss how a first-order EWPT can be induced by leptoquarks. We consider the coupling between the SM scalar doublet $H$ and an extra complex scalar leptoquark $S$ with a $SU(2)$ index $a$, which runs up to 1, 2 or 3 for singlet, doublet or triplet representations, respectively. In the simplest case, the scalar potential can be written as 
\begin{eqnarray}
V_0 = -\mu^2 |H|^2 + \lambda_H |H|^4 +\mu_S^2|S_a|^2 + \lambda_S |S_a|^4 + 2\lambda_{HS} |H|^2|S_a|^2
\end{eqnarray}
For simplicity, we only consider the minimal quartic interaction between Higgs and scalar leptoquark in the form of $|H|^2|S|^2$. Other forms of quartic interactions, such as $|H^\dagger S|^2$ for $SU(2)$ doublet leptoquark and $H^\dagger (\sigma^i S_i)(\sigma^j S_j)^\dagger H$ for $SU(2)$ triplet leptoquark, can lead to mass shifts between the $SU(2)$ components of leptoquarks after spontaneous symmetry breaking (SSB), as well as extra contributions to the thermal mass of the SM Higgs field. Such contributions can enhance the phase transition, but the effect can be taken into account effectively by shifting the minimal quartic coupling $\lambda_{HS}$. Focussing on the field $h$ in $H=(G^+,(h+iG^0)/\sqrt2)$ that becomes the SM Higgs boson after spontaneous symmetry breaking, the scalar potential reads
\begin{eqnarray}
V_0 = -\frac{\mu^2}{2} h^2 + \frac{\lambda_H}{4} h^4 +\frac{\mu_S^2}{2}\left(s_{a,1}^2+s_{a,2}^2\right) + \frac{\lambda_S}{4}\left(s_{a,1}^2+s_{a,2}^2\right)^2 + \frac{\lambda_{HS}}{2} h^2\left(s_{a,1}^2+s_{a,2}^2\right)
\end{eqnarray}
where $S_a=(s_{a,1}+i\,s_{a,2})/\sqrt2$. As the leptoquark is typically heavier than the electroweak scale, we assume $\mu_S^2 > 0$ in this research. Then the leptoquark mass after SSB is $m_S^2 = \mu_S^2 + \lambda_{HS} v_0^2$ with $v_0$ the standard model Higgs VEV. At tree level, the phase transition is second-order as the participation of $S$ does not vary the minimum of the scalar potential. However, by considering the finite temperature effective potential, the existence of leptoquarks modifies the minimum through the Higgs portal coupling at loop order. In this study, we consider the effective potential at one-loop level for simplicity, neglecting higher-order effects \cite{Niemi:2021qvp} that may vary the transition strength by 20\%. We also neglect renormalisation group corrections which have a smaller effect \cite{Schicho:2022wty}.

At one-loop level, the effective scalar potential receives contributions from zero-temperature correction $\Delta V_0^{\rm 1-loop}$ (Coleman-Weinberg potential) and finite-temperature correction $\Delta V_T^{\rm 1-loop}$ \cite{Quiros:1999jp}
\begin{eqnarray}
V_{\rm eff} (h,T)= V_0 + \Delta V_0^{\rm 1-loop}(h) + \Delta V_T^{\rm 1-loop}(h,T)\,.\label{eq:Veff}
\end{eqnarray}
The one-loop zero-temperature correction reads 
\begin{eqnarray}
\Delta V_0^{\rm 1-loop}(h) &=& \sum_{i\in b,f}  \frac{n_i}{64\pi^2} \left[ m_i^4(h) \left(\ln\frac{m_i^2(h)}{m_i^2(v_0)} -\frac32\right) + 2m_i^2(h)m_i^2(v_0)\right]\,,
\label{eq:V01L}
\end{eqnarray}
where $m_i^2=m_{0i}^2 +a_i h^2$ are the shifted masses with
\begin{subequations}
\begin{eqnarray}
&&m_{0\{t, W, Z, h, G, S\}}^2 = \{0, 0, 0, -\mu^2, -\mu^2, \mu_S^2\}\,,\\
&&a_{\{t, W, Z, h, G, S\}} = \{\frac{y_t^2}{2}, \frac{g^2}{4},\frac{g^2+{g'}^2}{4}, 3\lambda_H, \lambda_H, \lambda_{HS}\}\,,\\
&&n_{\{t, W, Z, h, G, S\}} = \{-12, 6, 3, 1, 3, n_S\}\,.
\end{eqnarray}
\end{subequations}
The quantity $v_0$ is the SM Higgs VEV at zero temperature. The degree of freedom $n_S$ in the complex $SU(3)$ triplet $S$, depending on the $SU(2)$ representation of the leptoquark, can be 6 for $SU(2)$ singlet, 12 for $SU(2)$ doublet or 18 for $SU(2)$ triplet. 

The one-loop finite-temperature correction in Eq.\ref{eq:Veff} is
\begin{eqnarray}
\Delta V_T^{\rm 1-loop}(h,T) &=& \sum_{i\in b} \frac{n_i T^4}{2\pi^2} J_b \left(\frac{m_i^2}{T^2}\right) + \sum_{i\in f} \frac{n_i T^4}{2\pi^2} J_f \left(\frac{m_i^2}{T^2}\right)\,
\end{eqnarray}
where $b$ and $f$ stand for bosons and fermions and 
\begin{eqnarray}
J_{b/f} \left(\frac{m_i^2}{T^2}\right) &=& \int_0^\infty dx x^2\ln\left[1 \mp e^{-\sqrt{x^2+m_i^2(h)/T^2}}\right]\,,
\end{eqnarray}
At high temperature $T \gtrsim m_i $, $J_b$ and $J_f$ can be expanded as
\begin{subequations}
\begin{eqnarray}
J_b \left(\frac{m_i^2}{T^2}\right) &\simeq& -\frac{\pi^4}{45} + \frac{\pi^2}{12}\frac{m_i^2}{T^2} - \frac{\pi}{6}\frac{m_i^3}{T^3} - \frac{1}{32}\frac{m_i^4}{T^4}\left(\ln\frac{m_i^2}{T^2} - c_b\right) +... \label{eq:Jb_app}\\
J_f \left(\frac{m_i^2}{T^2}\right) &\simeq&\frac{7\pi^4}{360} - \frac{\pi^2}{24}\frac{m_i^2}{T^2} - \frac{1}{32}\frac{m_i^4}{T^4}\left(\ln\frac{m_i^2}{T^2} - c_f\right) +...\label{eq:Jf_app}
\end{eqnarray}
\end{subequations}
with $c_b \simeq 5.4$ and $c_f \simeq 2.6$. At low temperature $T < m_i $, $J_b$ is exponentially suppressed as its argument increases.

The inclusion of Goldstone degrees of freedom in Eq.\eqref{eq:V01L} leads to an infrared logarithmic divergence as the Goldstone bosons are massless at the zero-temperature vacuum. Such a problem is also related to the perturbativity of couplings at high temperature \cite{Linde:1980ts} and can be solved by resuming the ring (daisy) diagrams \cite{Quiros:1999jp,Croon:2020cgk}. There are two different methods that are widely used for resummation. In the Parwani method \cite{Parwani:1991gq}, the shifted masses of scalars and the longitudinal models of the gauge bosons in the effective potential are replaced by the Debye masses $M_i^2(h,T)=m_i^2(h)+\Pi_i(T)$, where the self-energies $\Pi_i(T)$ are given by $\Pi_i (T)= b_i T^2$ with \cite{Curtin:2016urg}
\begin{eqnarray}
&&b_h = b_G = \frac{3g^2+{g'}^2}{16} + \frac{\lambda_H}{2} + \frac{y_t^2}{4}+ \frac{n_S \lambda_{HS}}{12} , \quad
b_{W} = b_{Z}(T) = \frac{11}{6} g^2 \,, \quad
b_{\gamma} = \frac{11}{6} {g'}^2  \,, \nonumber\\
&&b_S = 
\begin{dcases}
\frac{\lambda_{HS}}{3} + \frac{(n_S+2)\lambda_S}{12} + \frac34 g_3^2 + \frac14 Y^2 + \frac{{g'}^2}{16} & SU(2) \text{singlet} \,,\\
\frac{\lambda_{HS}}{3} + \frac{(n_S+2)\lambda_S}{12} + \frac34 g_3^2 + \frac14 Y^2 + \frac{3g^2+{g'}^2}{16} & SU(2) \text{ doublet and triplet.}
\end{dcases}
\nonumber\\
\end{eqnarray}
In the Arnold-Espinosa method \cite{Arnold:1992rz}, the replacement only happens in the mass cubic terms. While the Arnold-Espinosa resummation avoids some unphysical linear terms that shift the symmetric minimum away from the origin, it fails at low temperatures as the method relies on the expansion in Eq.\eqref{eq:Jb_app}. In this paper, we adapt the Parwani method in order to achieve a smooth connection to correct low-temperature thermal correction \cite{Cline:2011mm,Croon:2020cgk}. While the leptoquark coupling $Y$ is typically smaller than the unitarity \cite{Hiller:2017bzc}, the $SU(3)$ coupling can have significant contribution to the Debye mass of the leptoquarks. However, the contributions, not only from the $SU(3)$ coupling but also from other gauge couplings, play the same role as the one from self-interaction coupling $\lambda_S$ in the phase transition and thus can be absorbed effectively by shifting $\lambda_S$ to $\tilde{\lambda}_S$. This kind of shift can change the parameter space where first-order phase transition can happen. However, as $\lambda_S$ is unconstrained, the discussion on the effect of $\tilde{\lambda}_S$ is less meaningful unless when $\lambda_S$ approaches its perturbativity limit. Therefore we focus on the effect of varying the Higgs portal coupling $\lambda_{HS}$ and fix the value of $\tilde{\lambda}_S$ as in \cite{Curtin:2014jma,Beniwal:2017eik}. Considering the contributions from the gauge couplings and leptoquark-fermion couplings, we fix $\tilde{\lambda}_S$ to be 2.

When the phase transition happens at a low temperature, the effective potential can develop an imaginary part as the thermal masses of Goldstone bosons become negative. It has been pointed out in \cite{Weinberg:1987vp} that such an imaginary part remarks the decay rate of the quantum state minimising the Hamiltonian.

\begin{figure}[t!]
\begin{center}
\subfigure[\,$SU(2)$ singlet scalar leptoquark\label{fig:PT_S}]{\includegraphics[width=0.45\textwidth]{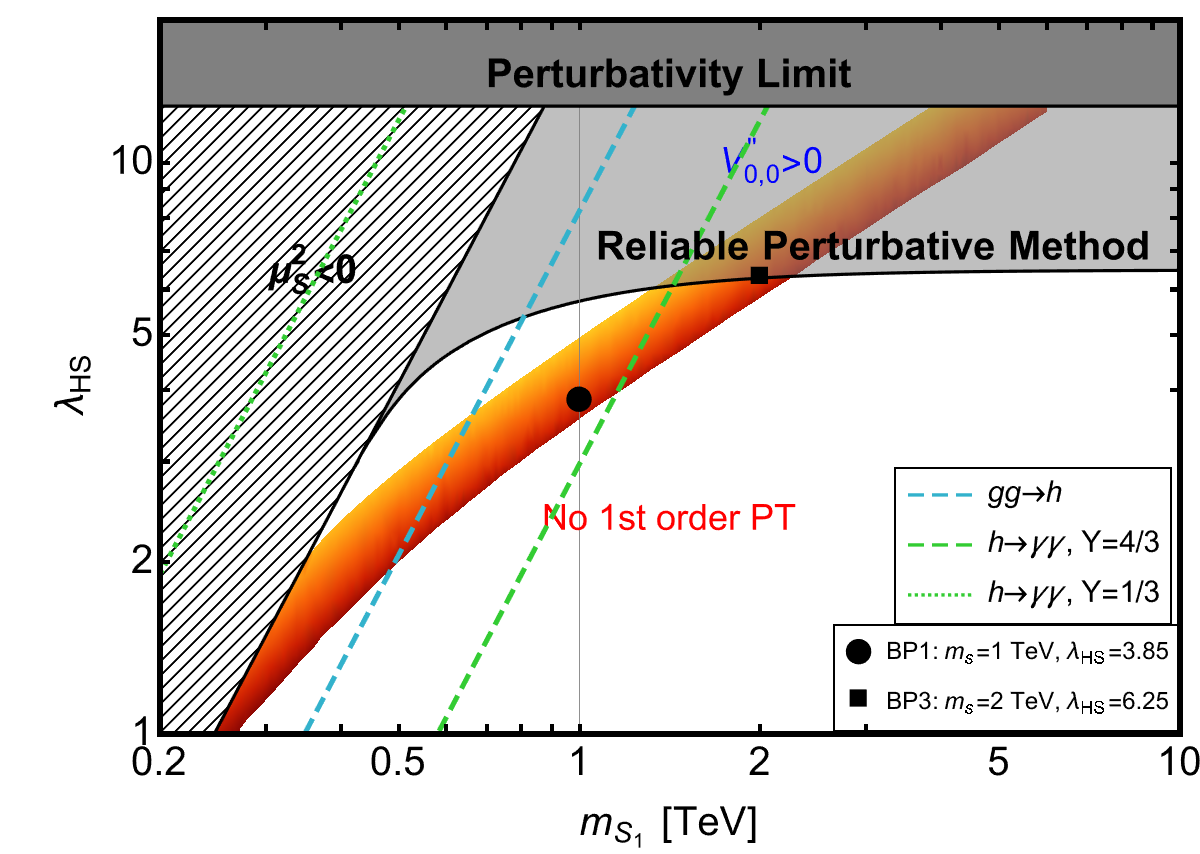}}\quad\quad
\subfigure[\,$SU(2)$ doublet scalar leptoquark\label{fig:PT_D}]{\includegraphics[width=0.45\textwidth]{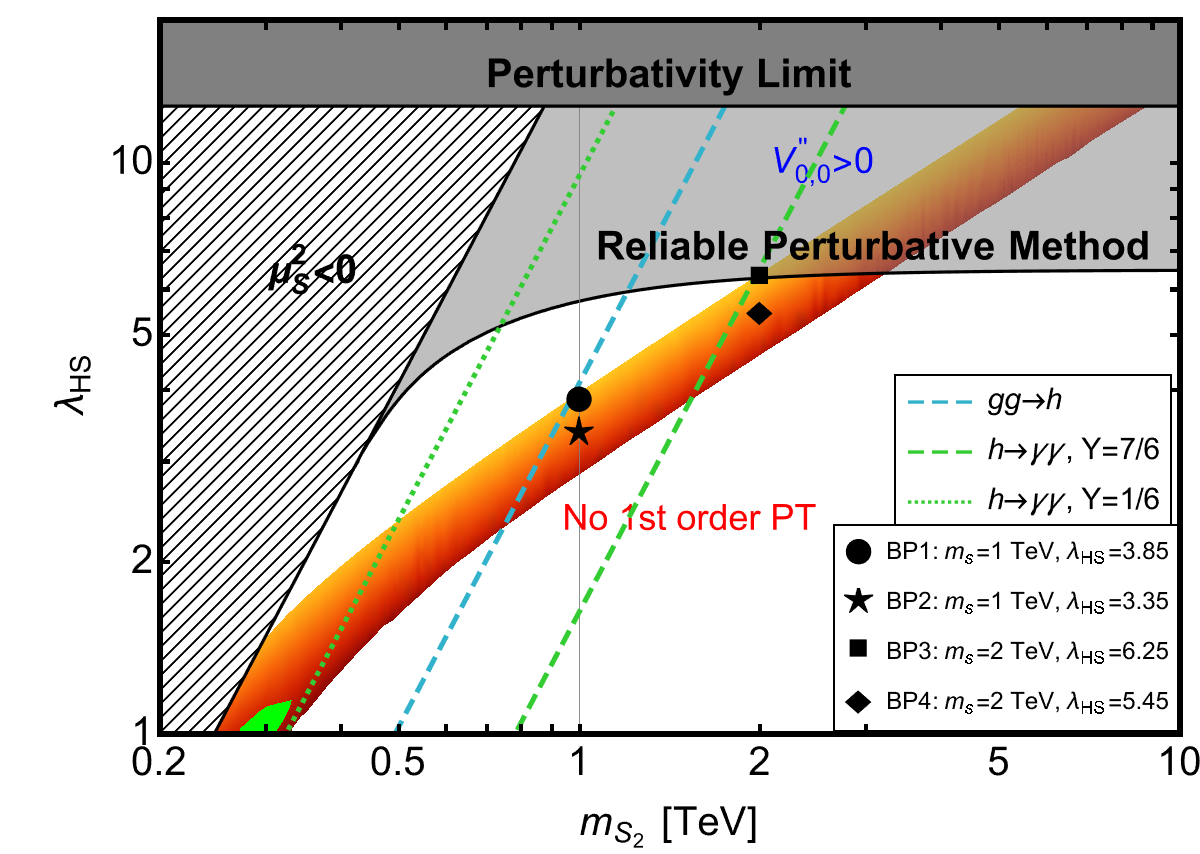}}
\subfigure[\,$SU(2)$ triplet scalar leptoquark\label{fig:PT_T}]{\includegraphics[width=0.45\textwidth]{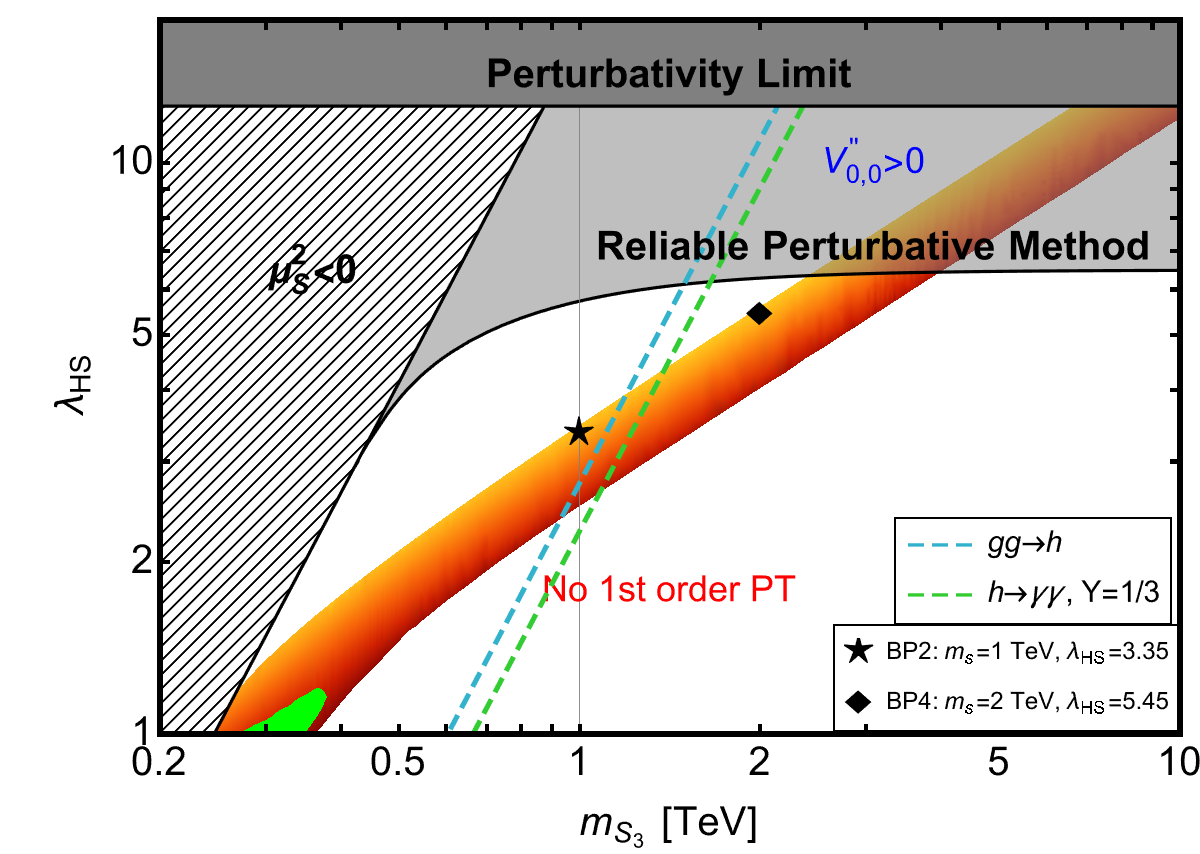}}\quad\quad
\subfigure[\,$SU(2)$ singlet, doublet and triplet scalar leptoquarks\label{fig:PT_all}]{\includegraphics[width=0.45\textwidth]{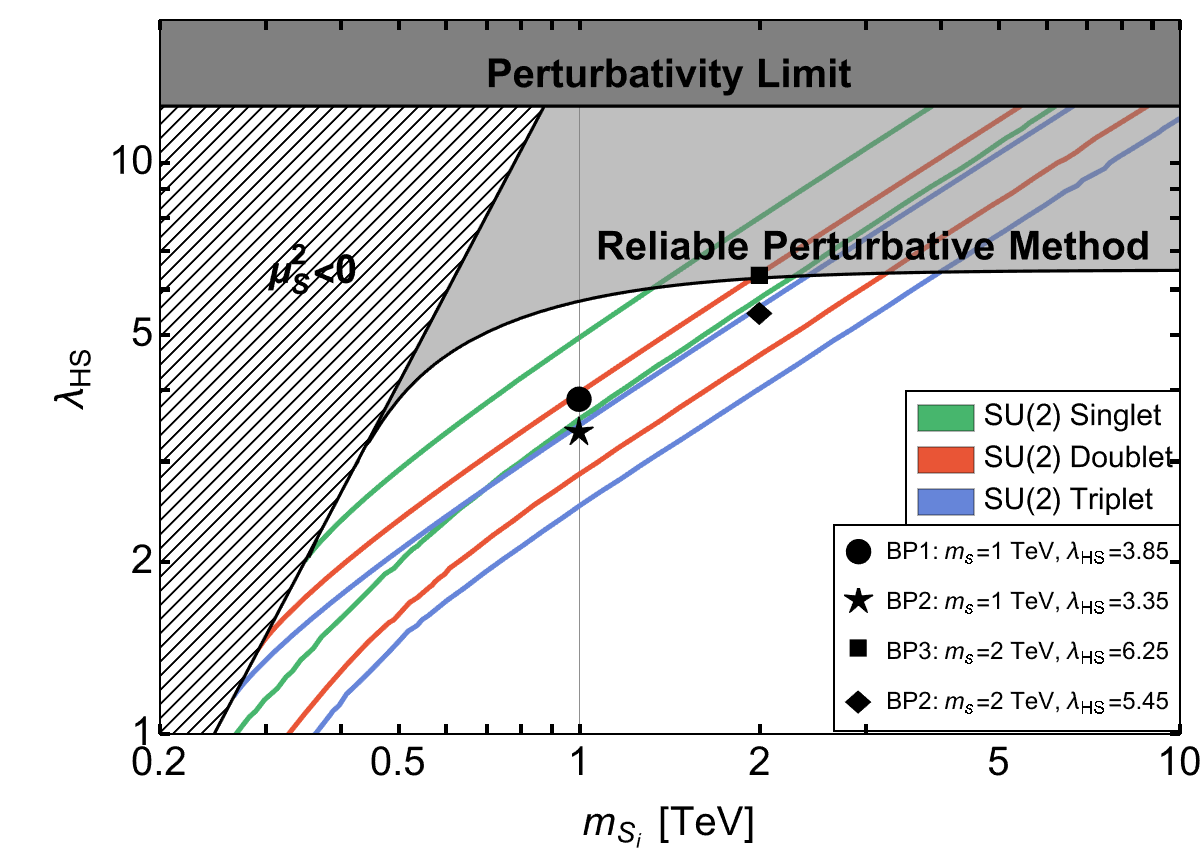}}
\includegraphics[width=0.41\textwidth]{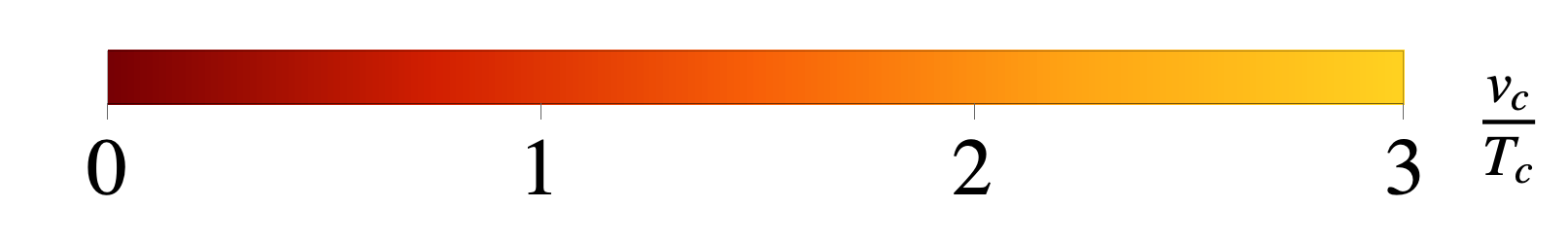}
\caption{\label{fig:PT} Allowed parameter space for first-order phase transition induced by different types of scalar leptoquark. Detailed discussion can be found in the text.}
\end{center}
\end{figure}

In the simplest case, the {\bf sufficient} \footnote{First-order phase transitions can happen even if the conditions listed below are not satisfied. An example is the green region in Fig.\ref{fig:PT_all}. However, we expect the strength of first order phase transition to be relatively weak in that region and unlikely to produce detectable gravitational waves. For simplicity, we only consider first-order phase transitions satisfying the conditions below.} conditions for an ``eligible'' first-order EWPT to occur are 
\begin{enumerate}
\item The electroweak minimum is the true minimum at zero temperature $T=0$ and $h=0$ is a local maximum ($V''(0,0)<0$).
\item At the temperature $T_2$ that $h=0$ change from a local maximum to a local minimum, there is another non-zero local minimum.
\end{enumerate}
The first condition ensures that the phase transition is completed today. If $h=0$ is a local minimum at zero temperature, the phase transition can only happen through tunnelling and the probability is too low to finish the transition to the true vacuum today. The second condition ensures that there are two minima existing simultaneously during the phase transition. In a scenario satisfying both conditions, the two minima of the scalar potential are degenerate at a critical temperature $T_c$. The allowed parameter spaces for first-order phase transition to happen are shown as the coloured regions in Fig.\ref{fig:PT}. The strength of the transition can be estimated by the ratio of the non-zero VEV and the critical temperature, $v_c/T_c$, which is shown are the colour in Fig.\ref{fig:PT}.  Above the coloured regions, the first-order EWPT is not eligible as condition 1 is not satisfied; below the coloured regions, first-order EWPT cannot happen because condition 2 is not satisfied. 

In Fig.\ref{fig:PT_S} to Fig.\ref{fig:PT_T}, the required coupling for first-order EWPT increases as the leptoquark becomes heavier in each $SU(2)$ representation of leptoquark. By comparing different panels and also by comparing the lines with different colours in Fig.\ref{fig:PT_all}, it can be figured out that the Higgs portal coupling required for first-order EWPT becomes smaller as the dimension of the leptoquark $SU(2)$ representation increases. Empirical expressions of the interesting parameter spaces can be obtained when the leptoquark is heavy. For leptoquark mass above 1 TeV, the allowed Higgs portal for eligible first-order EWPT to happen is roughly between $\{3.59,\,4.99\}\times (m_{S_1}/1\,\text{TeV})^{0.685}$ for singlet leptoquark, between $\{2.87,\,4.00\}\times (m_{S_2}/1\,\text{TeV})^{0.679}$ for doublet leptoquark and between $\{2.52,\,3.50\}\times (m_{S_3}/1\,\text{TeV})^{0.676}$ for triplet leptoquark.
 
A more complicated case can occur when the scalar potential develops two non-zero minima simultaneously after the temperature drops below $T_2$. In such a case, the transition has two steps: first to a non-zero minimum continuously through second-order phase transition and then to the larger non-zero minimum through first-order phase transition. The regions where such cases happen are marked as green in Fig.\ref{fig:PT_D} and Fig.\ref{fig:PT_T}. However, as the leptoquark is typically above 1 TeV, such regions are not of interest in this study. 

Since the coupling that is required for a first-order phase transition to appear grows as the scalar mass increases, the perturbative method used in the effective potential calculation can break down. To estimate the reliability of the perturbative method, we evaluate the loop contribution to the Higgs quartic coupling as has been done in \cite{Curtin:2014jma}. Since the main concern is the Higgs portal coupling $\lambda_{HS}$, we consider the higher loop contributions from the leptoquarks only involving Higgs portal vertices. Although the contribution to the Higgs quartic coupling from the leptoquarks is proportional to its internal degrees of freedom, the leptoquarks with different quantum numbers contribute separately as only the Higgs portal vertices are considered. Therefore, the ratio between the one-loop and two-loop corrections is independent of the leptoquark internal degrees of freedom and can be estimated as the one-loop correction from each leptoquark component
\begin{eqnarray}
\frac{\delta \lambda_H^{\rm2-loop}}{\delta \lambda_H^{\rm1-loop}} \sim \frac{\lambda_{HS}}{16\pi^2} \left(\ln\frac{\mu_S^2+\Lambda_{HS} h^2}{m_S^2} -\frac32\right)\,.
\end{eqnarray}
To ensure the reliability of the perturbative method, we choose $0.2$ to be the approximate threshold for the ratio above. In Fig.\ref{fig:PT}, the region where the perturbative method is no longer reliable is marked out by the shadows area.  

%To estimate the reliability of perturbative method, we use now power counting arguments to investigate contribution from higher loop diagrams to the effective potential. Since the main concern is the Higgs portal coupling $\lambda_{HS}$, we only consider the higher loop contributions from the leptoquarks. The resummation process helps us to take the higher order daisy diagrams into account. 

%%%%%%%%%%%%%%%%%%%%%%%%%%%%%%%%%%%%%%%%%%%%%%%%%%%%%%%%%%%%%%%%%%%%%
\subsection{Constraints on the Higgs portal coupling}
%%%%%%%%%%%%%%%%%%%%%%%%%%%%%%%%%%%%%%%%%%%%%%%%%%%%%%%%%%%%%%%%%%%%%

The new interaction between a scalar leptoquark and the Higgs doublet can affect the Higgs boson production and decay processes. The discrepancy between SM prediction and experimental measurement is commonly characterised by the $\varkappa$-factor, defined as $\varkappa_i=\sqrt{\Gamma_i^{\rm exp}/\Gamma_i^{\rm SM}}$ \cite{TheATLASandCMSCollaborations:2015bln,CMS:2015kwa}. The loop-induced contribution from leptoquark to the Higgs boson decay process into photons and the gluon-gluon production of Higgs boson are given by \cite{Dorsner:2016wpm}
\begin{eqnarray}
\varkappa_g &=& 1 + 0.24 \, \frac{\lambda_{HS}\,v^2}{m_S^2} N_S\\
\varkappa_\gamma &=& 1 - 0.052 \, \frac{\lambda_{HS}\,v^2}{m_S^2} N_c \sum_i Q_i^2
\end{eqnarray}
where the sum is taken over all $SU(2)$ components of the leptoquark and $Q_i$ is the electric charge of the $i$th component. $N_S$ is the number of the leptoquark $SU(2)$ components. The experimental measurements by the ATLAS collaboration are $\varkappa_g = 1.01^{+0.11}_{-0.09}$ and $\varkappa_\gamma = 1.02^{+0.08}_{-0.07}$ \cite{ATLAS:2022tnm}. Similar contribution appears in the decay channel of Higgs into a $Z$ boson and a photon as well, in the form of \cite{Dorsner:2016wpm}
\begin{eqnarray}
\varkappa_{Z\gamma} &=& 1 + 0.036\, \frac{\lambda_{HS}\,v^2}{m_S^2} N_c \sum_i Q_i \left(I_i^W - 0.23 Q_i\right)
\end{eqnarray}
where $I_i^W$ is the value of the weak isospin of the leptoquark. The value of $\varkappa_{Z\gamma}$ measured by CMS collaboration is $1.65^{+0.34}_{-0.37}$ \cite{CMS:2022dwd}. Despite abundant collider phenomena caused by the Higgs portal to leptoquarks, none of the observables can constrain the portal coupling restrictedly. When multiple leptoquarks appear in a model, the contributions from different types of leptoquarks can have opposite contributions to the $\varkappa$ parameters above. In order to visualise the effects of these observables, we consider the collider constraints under the assumption of a single leptoquark multiplet and show the maximal values of the Higgs portal allowed by $h\rightarrow\gamma\gamma$ and $gg\rightarrow h$ as the dashed and dotted lines in Fig.\ref{fig:PT_S} to Fig.\ref{fig:PT_T}. While the $gg\rightarrow h$ cross section is affected by the $SU(2)$ representation of the leptoquark, the $h\rightarrow\gamma\gamma$ cross section depends on the electric charge. For scalar leptoquark, assuming direct interaction to SM fermions, there are two different possible assignments of hypercharge for $SU(2)$ singlet and doublet and one assignment for $SU(2)$ triplet \cite{ParticleDataGroup:2020ssz}: $4/3$ or $1/3$ for singlet, $7/6$ or $1/6$ for doublet and $1/3$ for triplet. Although those constraints are currently weak, they are expected to be improved foreseeably by future experiments like HL-LHC \cite{CMS:2018qgz}, FCC \cite{dEnterria:2017dac,FCC:2018evy}, ILC \cite{Bambade:2019fyw} and CEPC \cite{An:2018dwb,Ruan:2021gap}. Moreover, the Higgs portal coupling also affects flavour violating processes like the $h\rightarrow\mu\tau$ or $\tau\rightarrow\mu\gamma$ decay which can be tested by precious measurements at colliders \cite{Dorsner:2016wpm,Crivellin:2020mjs}.

%%%%%%%%%%%%%%%%%%%%%%%%%%%%%%%%%%%%%%%%%%%%%%%%%%%%%%%%%%%%%%%%%%%%%
\section{Gravitational wave signals \label{sec:GW}}
%%%%%%%%%%%%%%%%%%%%%%%%%%%%%%%%%%%%%%%%%%%%%%%%%%%%%%%%%%%%%%%%%%%%%

During a first-order phase transition, the scalar field configuration tunnels from the zero vacuum to a non-zero vacuum locally in the form of bubbles. The scalar bubbles can then move, collide and expand. Sound waves and turbulence can be produced after the collision of bubbles. The gravitational wave can be produced through three different mechanisms \cite{Caprini:2015zlo,Weir:2017wfa}: {\bf collision} between the scalar bubbles, overlap of the {\bf sound wave} in the plasma and the fluid {\bf turbulence}. The total gravitational wave spectrum is the sum of the three contributions
\begin{eqnarray}
\Omega_{\rm tot}(f) = \Omega_{\rm coll}(f) + \Omega_{\rm sw}(f) + \Omega_{\rm turb}(f)\,.
\end{eqnarray}
All three contributions depend on the phase transition dynamics which is described by four key parameters: the wall velocity $v_w$, the inverse phase transition duration $\beta/\Hub_*$, the phase transition strength $\alpha_{T_*}$ and the transition temperature $T_*$. After these parameters are determined, the gravitational wave spectrum can be computed using results from numerical simulations. 

The crucial step in computing these key parameters is to compute the Euclidean action. To find the Euclidean action which is defined as the spacial integration of the effective Lagrangian, a solution of the Euclidean equation of motion is required, which is generally not solvable analytically. For further details see Appendix \ref{app:GWinPT}. A common treatment for particles of electroweak scale or below is to make an approximation using Eq.\eqref{eq:Jb_app} and Eq.\eqref{eq:Jf_app} after which the effective potential can be simplified into a quartic function of the scalar field and a semi-analytical expression for the Euclidean action can be derived \cite{Dine:1992wr,DiBari:2021dri}. However, as the leptoquark is typically above TeV scale \cite{CMS:2020gru,CMS:2020wzx,ATLAS:2021oiz,ATLAS:2022fho}, the one-loop finite-temperature correction from leptoquark is exponentially suppressed and thus negligible. On the other hand, the approximation in Eq.\eqref{eq:Jb_app} and Eq.\eqref{eq:Jf_app} are no longer eligible in the parameter space of interest. Therefore the Euclidean equation of motion is solved numerically in this work.

\begin{figure}[t!]
\begin{center}
\subfigure[\,$SU(2)$ singlet leptoquark $m_S=1$ TeV, $v_*/T_*=1.16$]{\includegraphics[width=0.45\textwidth]{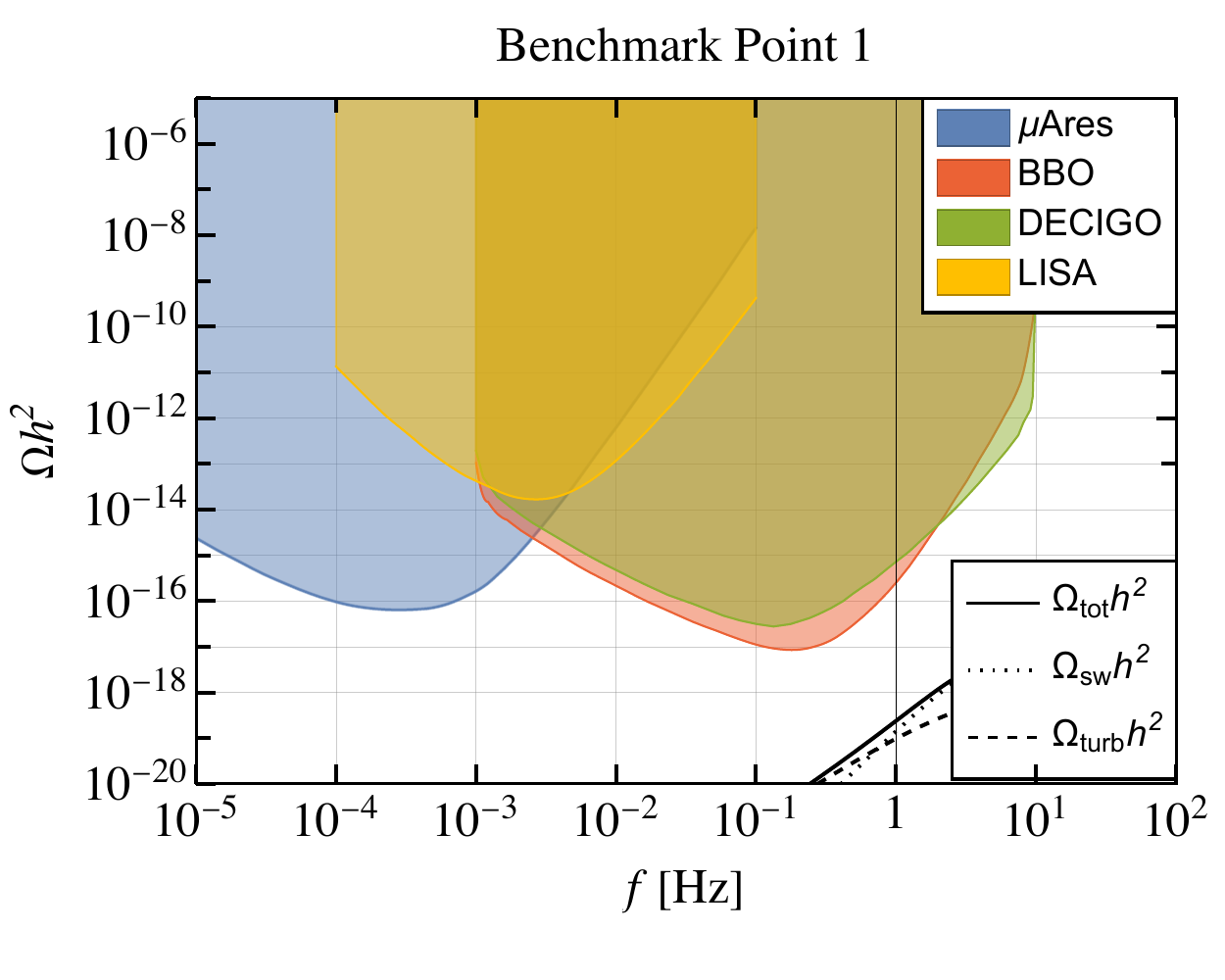}}\quad\quad
\subfigure[\,$SU(2)$ doublet leptoquark $m_S=1$ TeV, $v_*/T_*=3.97$]{\includegraphics[width=0.45\textwidth]{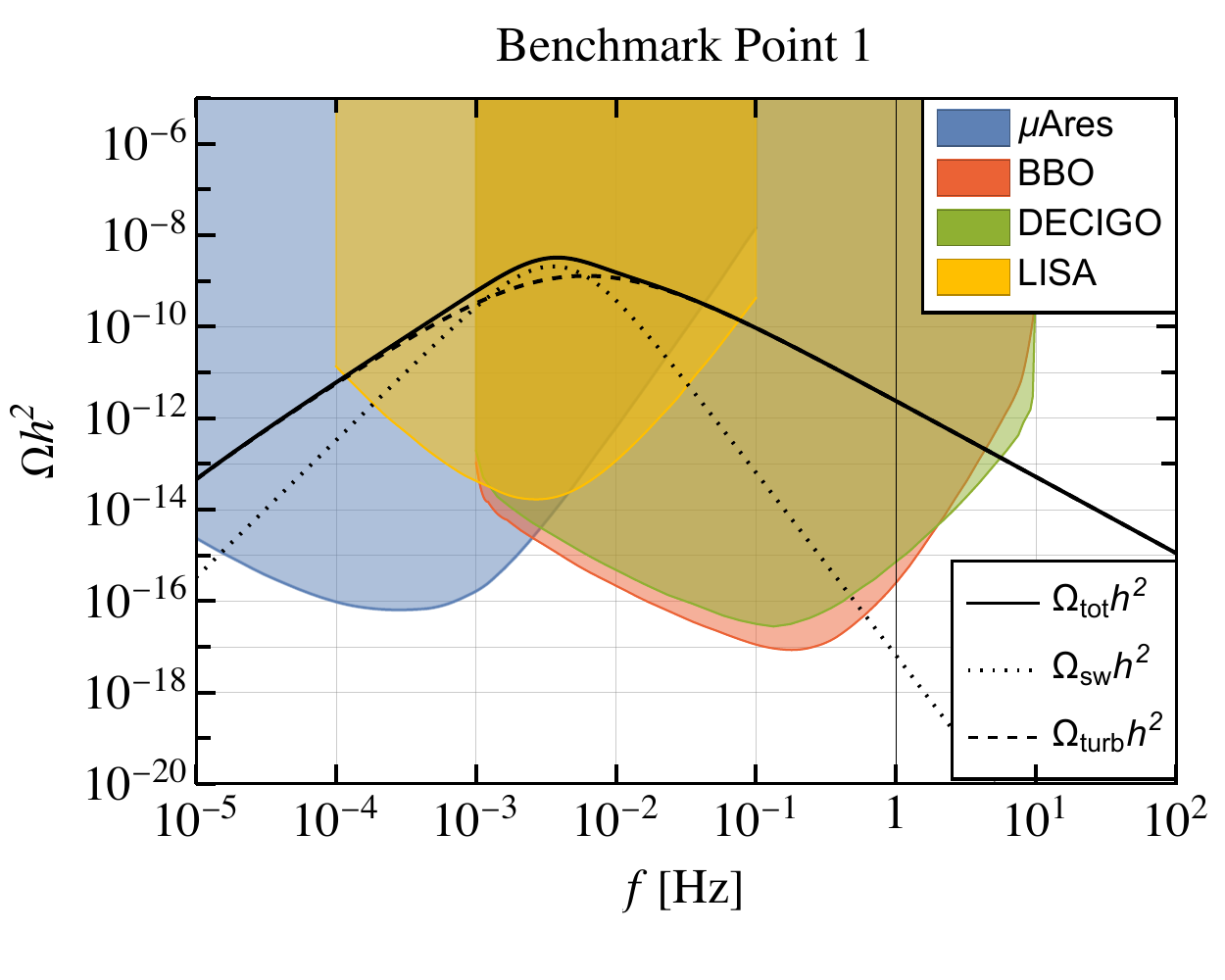}}
\subfigure[\,$SU(2)$ doublet leptoquark $m_S=1$ TeV, $v_*/T_*=1.75$]{\includegraphics[width=0.45\textwidth]{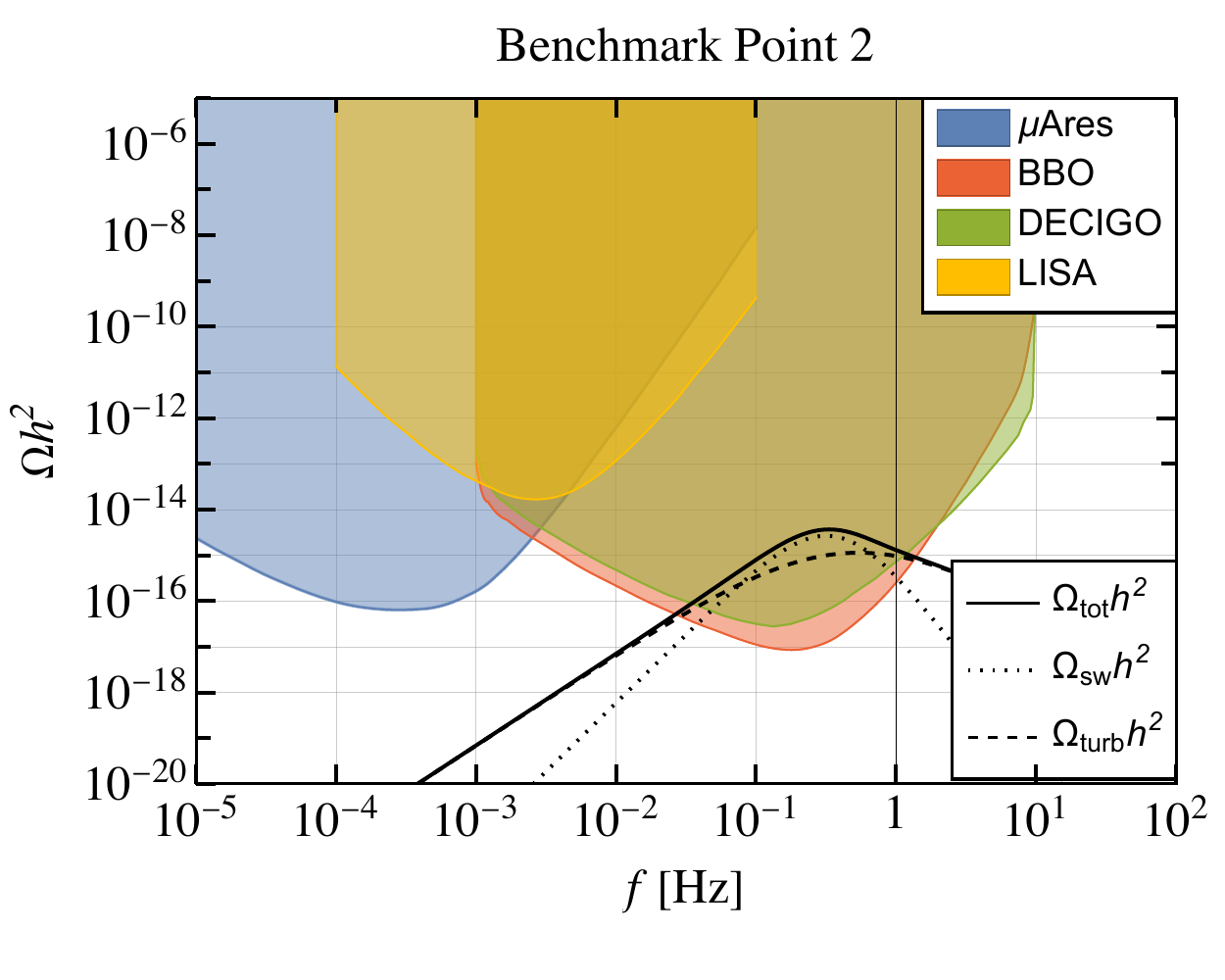}}\quad\quad
\subfigure[\,$SU(2)$ triplet leptoquark $m_S=1$ TeV, $v_*/T_*=3.42$]{\includegraphics[width=0.45\textwidth]{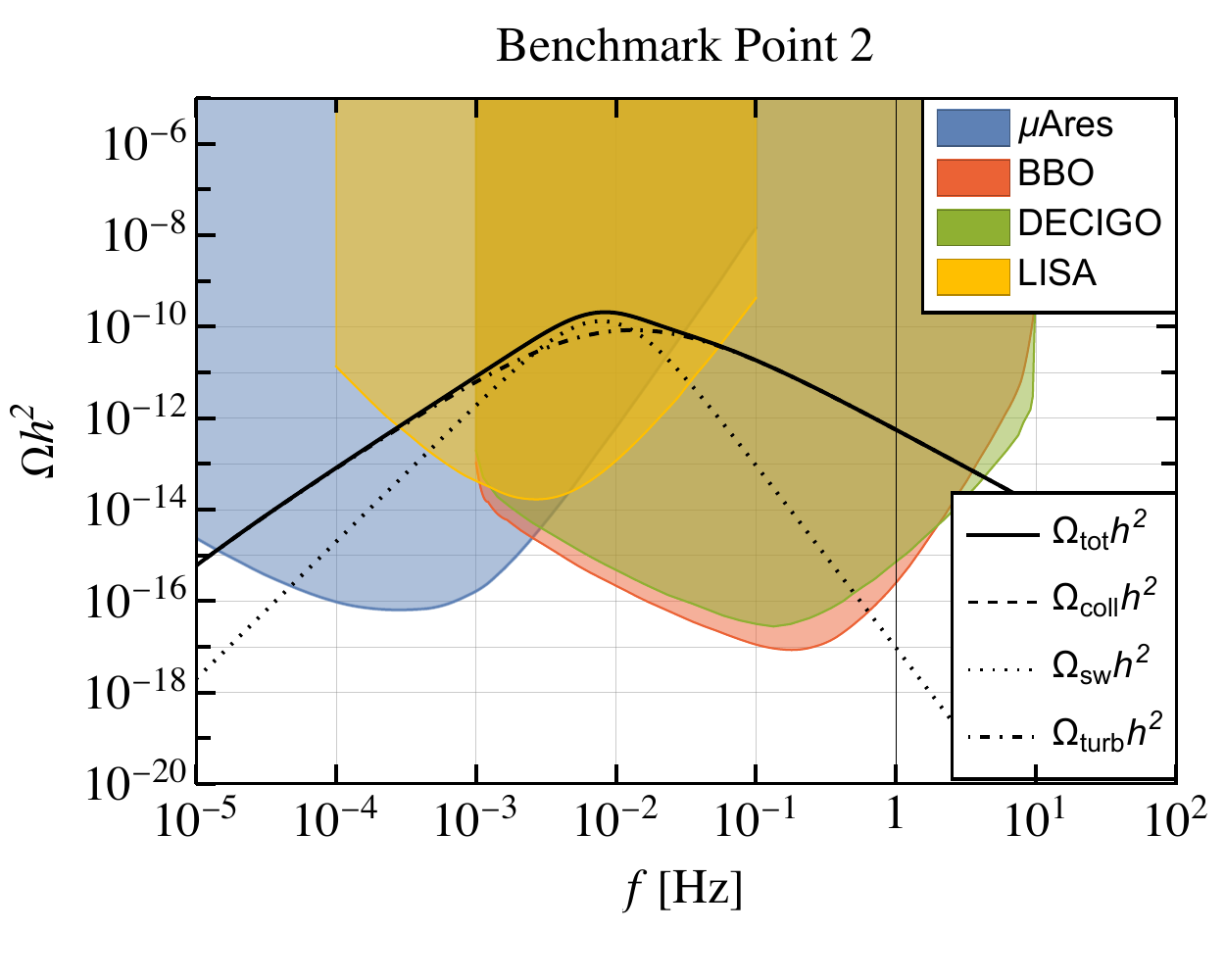}}
\caption{\label{fig:fix_BP1} Gravitational wave signals for benchmark cases with 1 TeV leptoquarks. See the text for further discussion and explanation.}
\end{center}
\end{figure}

\begin{figure}[t!]
\begin{center}
\subfigure[\,$SU(2)$ singlet leptoquark $m_S=2$ TeV, $v_*/T_*=1.22$]{\includegraphics[width=0.45\textwidth]{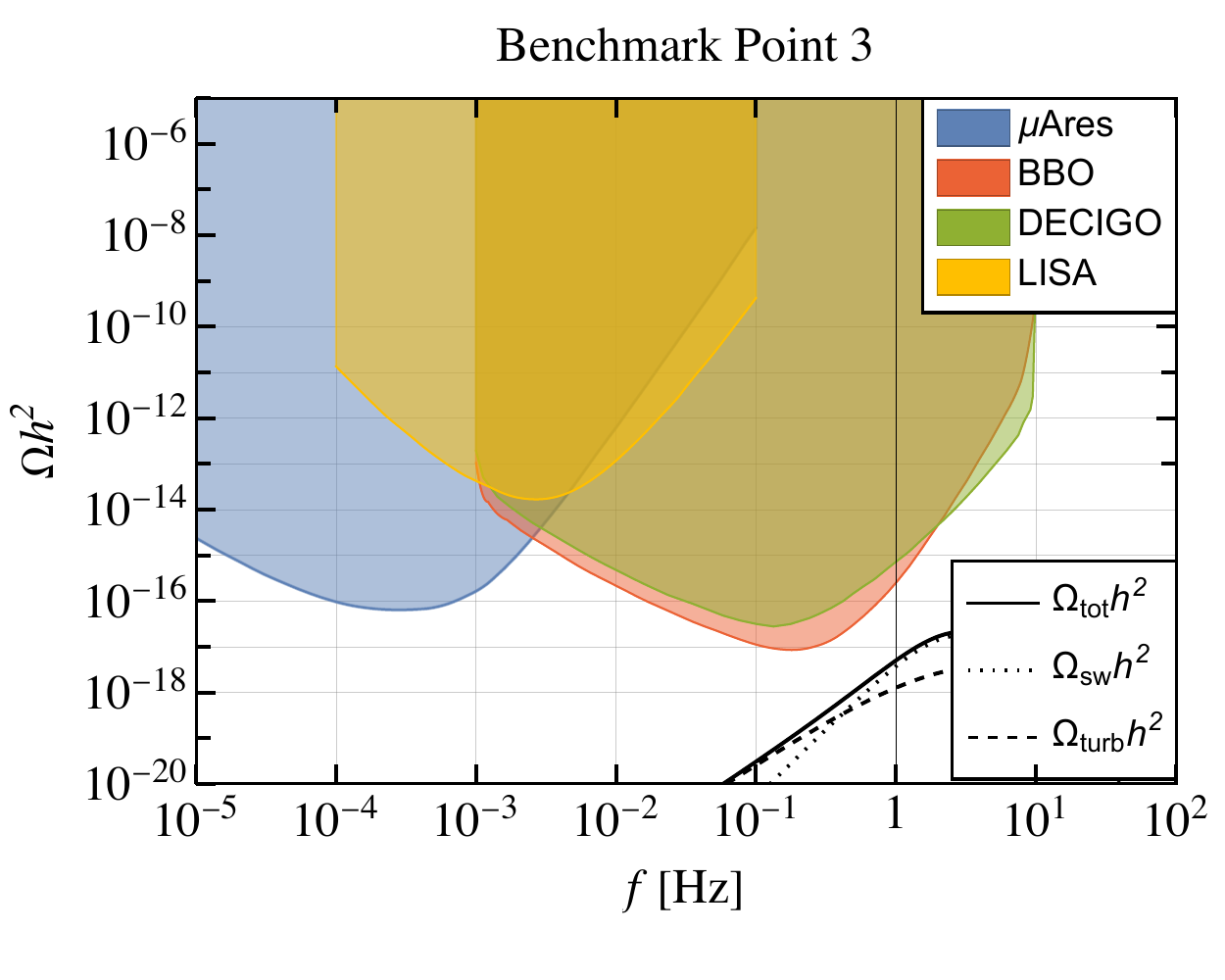}}\quad\quad
\subfigure[\,$SU(2)$ doublet leptoquark $m_S=2$ TeV, $v_*/T_*=4.78$]{\includegraphics[width=0.45\textwidth]{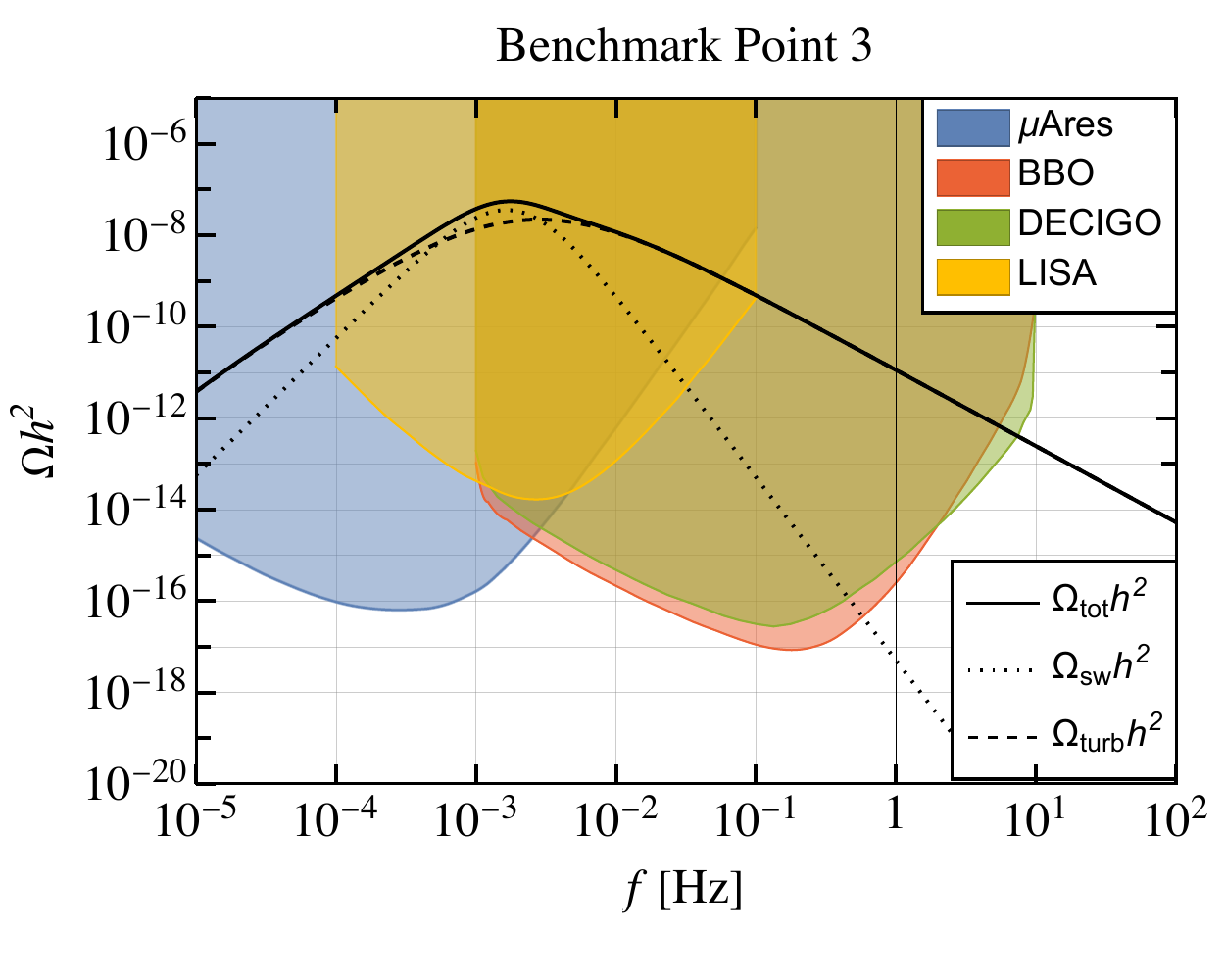}}
\subfigure[\,$SU(2)$ doublet leptoquark $m_S=2$ TeV, $v_*/T_*=1.87$]{\includegraphics[width=0.45\textwidth]{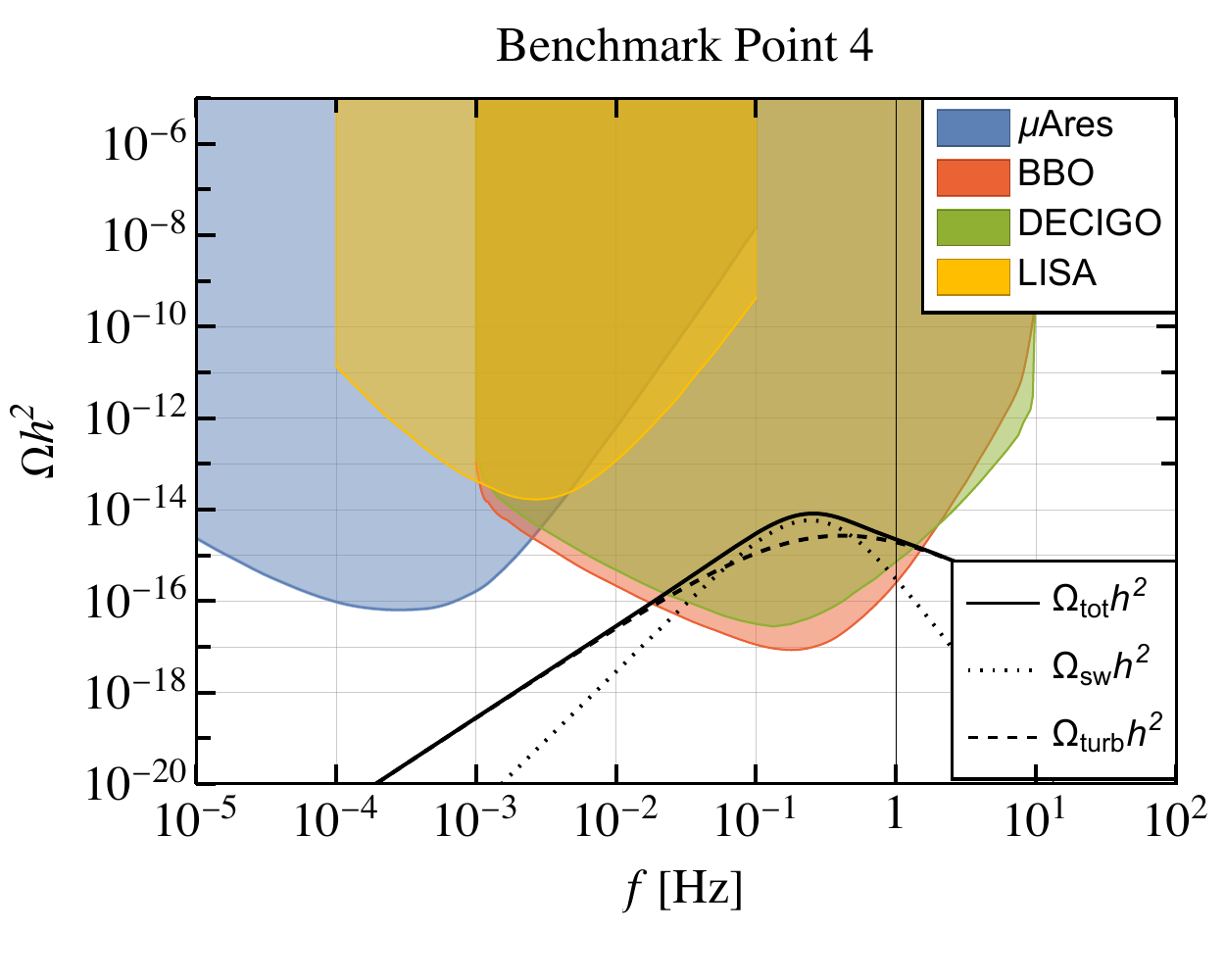}}\quad\quad
\subfigure[\,$SU(2)$ triplet leptoquark $m_S=2$ TeV, $v_*/T_*=4.38$]{\includegraphics[width=0.45\textwidth]{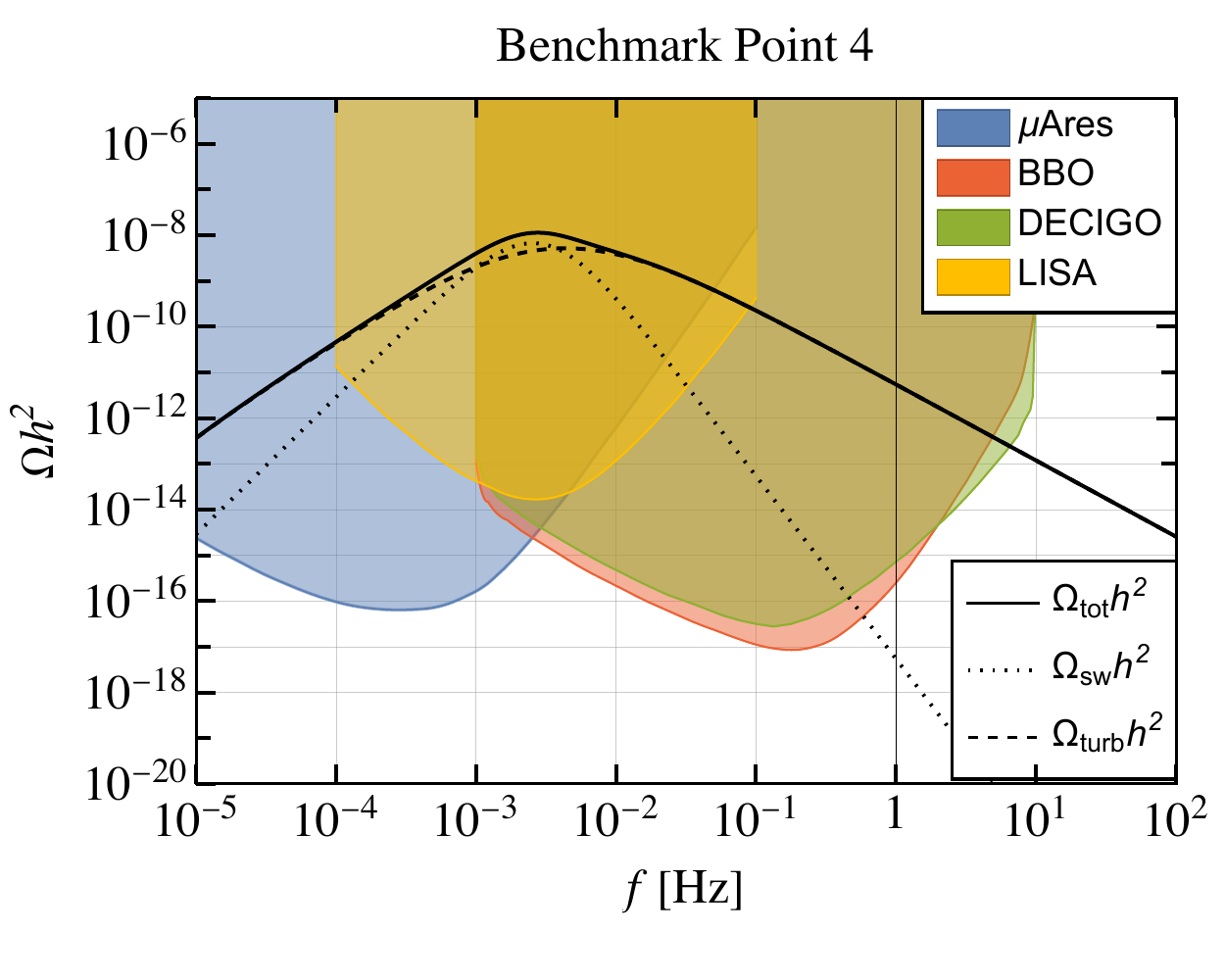}}
\caption{\label{fig:fix_BP2} Gravitational wave signals for benchmark cases with 2 TeV leptoquarks. See the text for further discussion and explanation.}
\end{center}
\end{figure}

In Fig.\ref{fig:fix_BP1} and Fig.\ref{fig:fix_BP2}, we show the gravitational wave produced from first-order EWPT for the 4 benchmark cases in Fig.\ref{fig:PT}. The cases with 1 TeV leptoquarks are shown in Fig.\ref{fig:fix_BP1} and those with 1 TeV leptoquarks are shown in Fig.\ref{fig:fix_BP2}. In order to compare the result with detections, we shadow the region that can be detected by BBO \cite{Corbin:2005ny}, DECIGO \cite{Kawamura:2019jqt,Kawamura:2020pcg}, LISA \cite{Caprini:2015zlo} and $\mu$Ares \cite{Sesana:2019vho} with different colours. The gravitational waves produced by different sources during the phase transition are also shown independently. As the average bubble radius $R_*$ at the percolation temperature is much larger than the initial bubble radius $R_0$, the gravitational wave produced by bubble collision is subdominant and can be neglected. We have also found that, for all the cases, the gravitational wave is dominantly produced by the turbulence at most frequencies, except around the peak frequency for the gravitational wave produced by the sound wave. The reason is that the sound wave period $\tau_{sw}$ is relatively small compared with the Hubble time, which means more energy budget in the fluid motion is released in the form of turbulence than the sound wave. However, the calculation of gravitational wave from turbulence after a phase transition has a relatively large uncertainty, and the assumption of full conversion of the fluid motion energy into turbulence can also lead to an overestimation of the gravitational wave strength. Therefore the contribution from turbulence is shown as a reference on its upper bound. As the gravitational wave produced by sound wave dominates when it peaks, the sensitivity to Higgs portal coupling is not affected.

\begin{figure}[t!]
\begin{center}
\includegraphics[width=0.45\textwidth]{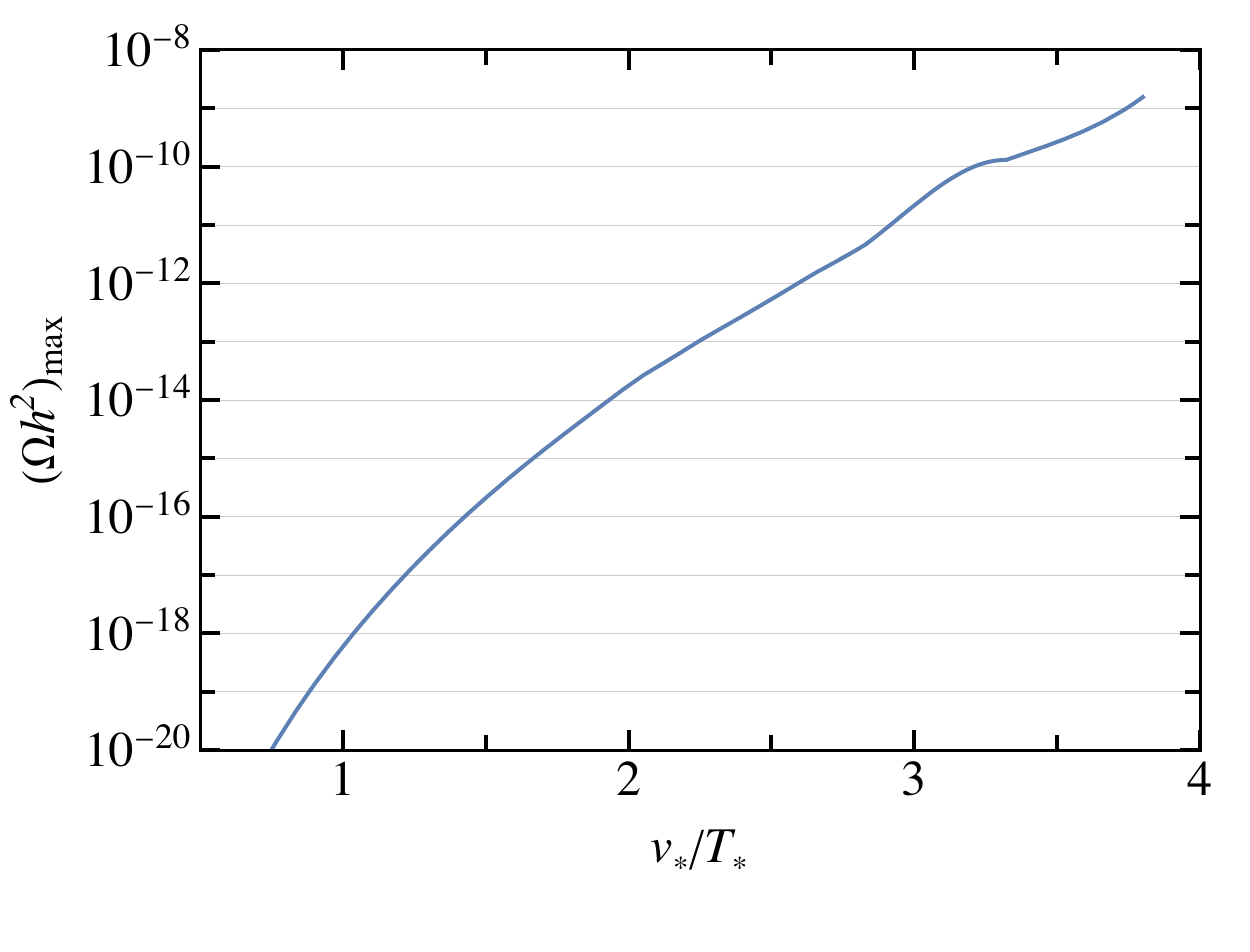}
\includegraphics[width=0.45\textwidth]{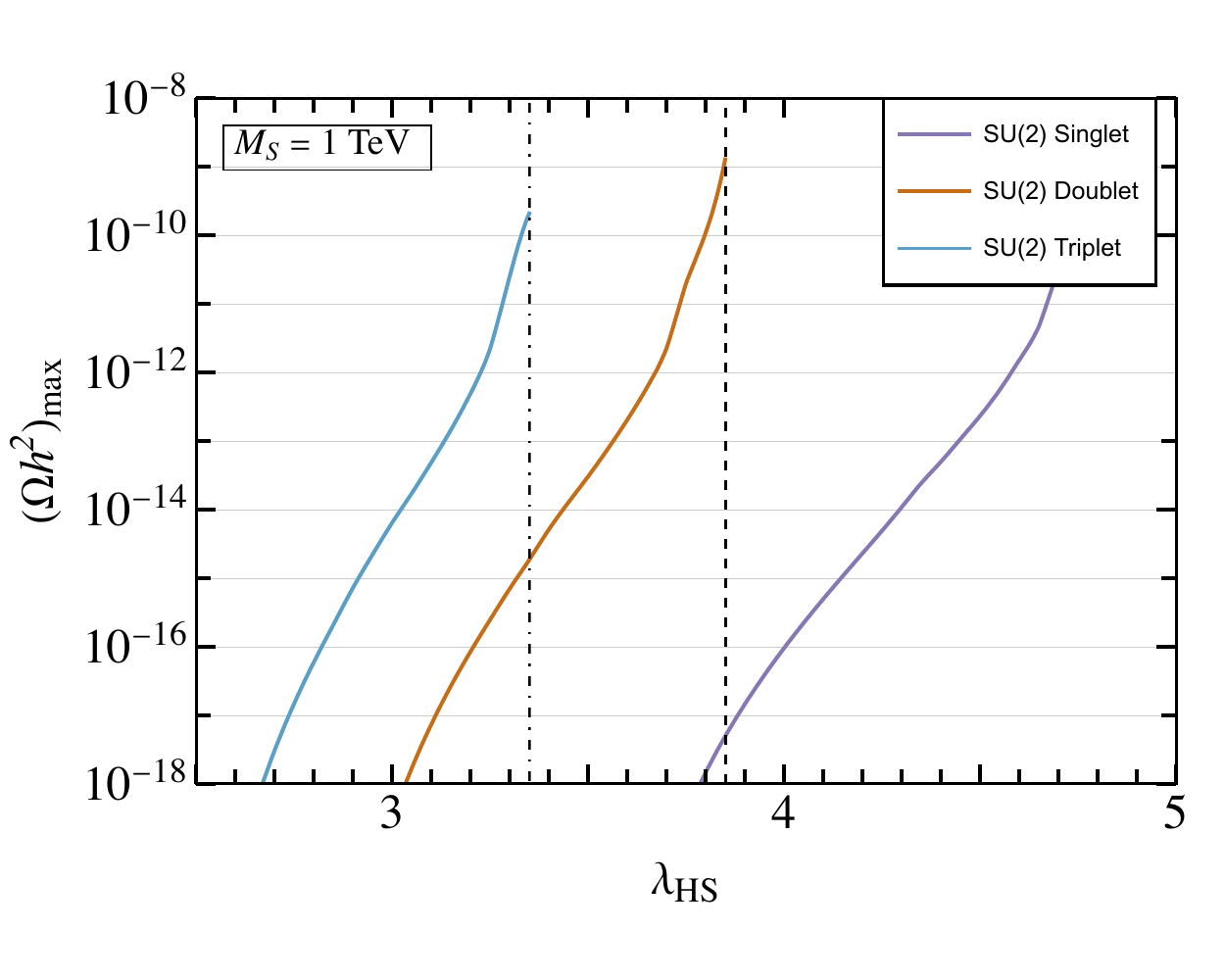}
\caption{\label{fig:peak_strength} Left panel: Maximal strength of the gravitational wave produced as a function of transition strength $v_*/T_*$. Right panel: Maximal strength of gravitational wave produced by first-order EWPT induced by different type of leptoquarks of 1 TeV as a function of the Higgs portal coupling $\lambda_{HS}$. See the text for further discussion and explanation.}
\end{center}
\end{figure}

By comparing the panels, it can be noticed that the gravitational wave produced from first-order EWPT relies on the strength of the transition. To illustrate the relation more explicitly, we show the dependence of gravitational wave signal peak values on the strength of the phase transition in the left panel of Fig.\ref{fig:peak_strength}. Here, instead of $v_c/T_c$ in the previous section, the strength of transition is estimated by the ratio of the non-zero minimum of the scalar potential and temperature when the phase transition happens, i.e. when the probability of bubble nucleation is significant. The temperature $T_*$ is defined by the temperature when one bubble is nucleated per unit volume per unit time and the non-zero VEV at $T_*$ is denoted as $v_*$. We find that the gravitational wave is testable when $v_*/T_*$ is roughly larger than $1.22$. In the right panel of Fig.\ref{fig:peak_strength}, we show how the gravitational wave signal peak values rely on the Higgs portal coupling when the masses of leptoquarks are 1 TeV. For the first-order phase transitions induced by different types of leptoquarks, the gravitation wave signals achieve the same peak value when the Higgs portal couplings are in different ranges. The dashed and dot-dashed lines corresponds to the benchmark cases 1 and 2, respectively.

For the same benchmark point, the gravitational wave produced during first-order EWPT induced by leptoquark with a smaller intrinsic degree of freedom is stronger. In both Fig.\ref{fig:fix_BP1} and Fig.\ref{fig:fix_BP2}, (a) and (b) show the result for the same benchmark point but in the case of singlet and doublet leptoquark, respectively, while (c) and (d) show the result for the same benchmark point but in the case of singlet and doublet leptoquark, respectively. It is clear that for the same coupling, the first-order EWPT induced by the $SU(2)$ multiplet with a higher dimension produces stronger gravitational waves. We also mark out the two benchmark values of the portal coupling in the right panel of Fig.\ref{fig:peak_strength} as the vertical lines. Supposing that the Higgs portal is measured to be in the region where first-order phase transition appears by future collider experiments, the gravitational waves detection provides an alternative method to further test the Higgs portal as well as determine the $SU(2)$ representation of leptoquarks.

%%%%%%%%%%%%%%%%%%%%%%%%%%%%%%%%%%%%%%%%%%%%%%%%%%%%%%%%%%%%%%%%%%%%%
\section{Conclusion\label{sec:con}}
%%%%%%%%%%%%%%%%%%%%%%%%%%%%%%%%%%%%%%%%%%%%%%%%%%%%%%%%%%%%%%%%%%%%%

In this paper, we have explored the possibility that first-order EWPT induced by the coupling between a scalar leptoquark and the SM Higgs boson produces detectable gravitational wave signals. We have considered different $SU(2)$ representations of the scalar leptoquark, including singlet, doublet and triplet. Despite the lack of VEV for leptoquark itself, a first-order EWPT can be induced due to the 1-loop order effects. In general, with first-order EWPTs, gravitational waves can be produced by multiple processes in the dynamical evolution of the scalar bubbles. The resulting gravitational waves form a stochastic background that can be probed by gravitational wave detectors.

We have calculated the effective potential of the SM Higgs field in the presence of a scalar leptoquark,
including tree level and 1-loop level contributions as well as the resummation over the ring/daisy diagrams. By applying the conditions for first-order EWPT, we have found that the leptoquark can induce a first-order EWPT in the parameter space allowed by collider constraints and can be tested by future Higgs precision experiments. Enhanced by the internal degree of freedom of the particular leptoquark, we found that the leptoquark in the $SU(2)$ representation with a higher dimension requires smaller coupling in order to trigger a first-order EWPT. 

We have followed the standard procedure to compute the gravitational wave spectrum during first-order EWPTs. It turns out that the gravitational wave spectrum is mainly determined by the strength of the phase transition characterised by the ratio of the non-zero VEV and the temperature at the time that the transition happens. However, due to the difference in internal degrees of freedom, the strengths of first-order EWPTs induced by leptoquarks with the same masses and Higgs portal couplings but different $SU(2)$ nature are different. Since the gravitational wave signals differ, this provides a possibility to determine the $SU(2)$ representation of the leptoquarks through the observations of gravitational wave in particular regions of parameter space.

%%%%%%%%%%%%%%%%%%%%%%%%%%%%%%%%%%%%%%%%%%%%%%%%%%%%%%%%%%%%%%%%%%%%%
\appendix
%%%%%%%%%%%%%%%%%%%%%%%%%%%%%%%%%%%%%%%%%%%%%%%%%%%%%%%%%%%%%%%%%%%%%

%%%%%%%%%%%%%%%%%%%%%%%%%%%%%%%%%%%%%%%%%%%%%%%%%%%%%%%%%%%%%%%%%%%%%
\section{Production of gravitational waves during a first-order phase transition \label{app:GWinPT}}
%%%%%%%%%%%%%%%%%%%%%%%%%%%%%%%%%%%%%%%%%%%%%%%%%%%%%%%%%%%%%%%%%%%%%

As a beginning to discuss the phase transition dynamics, we need to determine the percolation temperature $T_*$. We use the condition derived in \cite{Ellis:2018mja} to determine the percolation temperature, which is
\begin{eqnarray}
I(T)\equiv\frac{4\pi}{3} \int_T^{T_c}\frac{\Gamma(T')}{T'^4\Hub(T')}\left(\int_T^{T'}\frac{dT''}{\Hub(T'')}\right)^3dT' =0.34\,,
\end{eqnarray}
The bubble nucleation rate $\Gamma$ is given by \cite{Ellis:2020nnr}
\begin{eqnarray}
\Gamma(T)=T^4\left(\frac{S_E}{2\pi}\right)^{3/2}e^{-S_E}\,,
\end{eqnarray}
where $S_E$ is the Euclidean action. At finite temperature, the four-dimensional euclidean action $S_E$ can be directly related to the three-dimensional Euclidean action $S_3$ by the relation $S_E = S_3/T$. With the $O(3)$ symmetry at high temperature, $S_3$ is defined as
\begin{eqnarray}
S_3 = 4\pi \int_0^\infty s^2 \left[ \frac12\left(\frac{dh}{ds}\right)^2 + V_{\rm eff}(h)\right]ds \,,
\end{eqnarray}
By solving the corresponding Euclidean equation of motion
\begin{eqnarray}
\frac{d^2h}{ds^2} + \frac{2}{s}\frac{dh}{ds} - \frac{dV_{\rm eff}}{dh} = 0 \,
\end{eqnarray}
with boundary conditions
\begin{eqnarray}
\left.\frac{dh}{ds}\right|_{s=0} = 0 \,, \quad \text{and}\quad\,\lim_{s\rightarrow\infty}h(s) = 0\,,
\end{eqnarray}
one can obtain the Euclidean action as a function of the temperature.

Given $T_*$ and $S_E$, the inverse phase transition duration can be expressed as 
\begin{eqnarray}
\frac{\beta}{\Hub_*}=T \left.\frac{d S_E (T)}{dT}\right|_{T=T_*}\,,
\end{eqnarray}
where $\Hub_*$ is the Hubble parameter at $T_*$. The phase transition strength $\alpha$ is the ratio of the difference in the trace anomaly $\theta$ to the radiation energy density at the transition temperature, i.e. $\alpha_{T_*}=\Delta\theta(T_*)/\rho(T_*)$. The difference in the trace anomaly is given by
\begin{eqnarray}
\Delta\theta(T)=- \Delta V + \frac{T}{4} \frac{d}{dT}\Delta V\,,\quad \Delta V=V_{\rm eff}^{\slashed{0}}(T)-V_{\rm eff}(0,T)
\end{eqnarray}
with $V_{\rm eff}^{\slashed{0}}(T)$ the effective potential at the non-zero minimum. 
In the thin-shell limit, the contributions to the GW spectrum from different sources are given by \cite{Ellis:2020nnr}
\begin{subequations}
\begin{eqnarray}
\Omega_{\rm coll}(f) &=& 2.30\times10^{-3}\left(R_*\Hub_*\right)^2 \left(\frac{\kappa_{\rm coll} \alpha_{T_*}}{1+\alpha_{T_*}}\right)^2 S_{\rm coll}(f)\,,\\
\Omega_{\rm sw}(f) &=& 0.384 \left(\tau_{\rm sw}\Hub_*\right) \left(R_*\Hub_*\right) \left(\frac{\kappa_{\rm sw} \alpha_{T_*}}{1+\alpha_{T_*}}\right)^2 S_{\rm sw}(f)\,,\\
\Omega_{\rm turb}(f) &=&6.85 \left(1- \tau_{\rm sw}\Hub_*\right) \left(R_*\Hub_*\right)\left(\frac{\kappa_{\rm sw} \alpha_{T_*}}{1+\alpha_{T_*}}\right)^{\frac32} S_{\rm turb}(f)\,.
\end{eqnarray}
\end{subequations}
where $R_*$ is the average bubble radius that can be estimated by \cite{Caprini:2019egz}
\begin{eqnarray}
R_*\Hub_*\simeq(8\pi)^{1/3}\text{Max}(v_w,\,c_s)\frac{\Hub_*}{\beta}\,.
\end{eqnarray}
$c_s$ is the sound speed in the plasma and $v_w$ is the wall speed that can be estimated analytically as \cite{Ellis:2022lft,Lewicki:2021pgr}
\begin{eqnarray}
v_w\simeq
\begin{dcases}
\sqrt{\frac{\Delta V}{\Delta\theta}}\,, & \sqrt{\frac{\Delta V}{\Delta\theta}}<v_J\,,\\
1\,, & \sqrt{\frac{\Delta V}{\Delta\theta}} > v_J\,,
\end{dcases}\,\quad\quad
v_J=\frac{1}{\sqrt{3}}\frac{1+\sqrt{3\alpha^2+2\alpha}}{1+\alpha}\,.
\end{eqnarray}
$\kappa_{\rm coll}$, $\kappa_{\rm sw}$ and $\kappa_t$ are the efficiency factors characterising the energy conversion during the phase transition. The efficiency factor for bubble collision is defined as the ratio of the bubble wall energy and the total released energy, i.e. $\kappa_{\rm coll}=E_{\rm wall}/E_V$

\begin{eqnarray}
\kappa_{\rm coll} &=& 
\begin{dcases}
\left[1-\frac13 \left(\frac{\tilde{\gamma}_*}{\gamma_{\rm eq}}\right)^2\right]\left(1-\frac{\alpha_\infty}{\alpha}\right)\,, & \tilde{\gamma}_*<\gamma_{\rm eq}\,,\\
\frac23\frac{\tilde{\gamma}_*}{\gamma_{\rm eq}}\left(1-\frac{\alpha_\infty}{\alpha}\right)\,, & \tilde{\gamma}_* > \gamma_{\rm eq}\,,
\end{dcases}
\end{eqnarray}
where $\gamma_{\rm eq}$ is the bubble wall Lorentz factor when the bubble is in equilibrium and $\tilde{\gamma}_*$ is the bubble wall Lorentz factor for a bubble with radius $R_*$. $\alpha_\infty$ is defined as the ratio of the contribution from $1\rightarrow1$ transitions to the pressure difference to the radiation energy density. For more details see \cite{Ellis:2019oqb,Ellis:2020nnr}. The efficiency factor for GW production from sound waves is given by
\begin{eqnarray}
\kappa_{\rm sw} &=& \frac{\alpha_{\rm eff}}{\alpha} \frac{\alpha_{\rm eff}}{0.73 + 0.083\sqrt{\alpha_{\rm eff}} + \alpha_{\rm eff}}\,  \quad\text{with}\quad 
\alpha_{\rm eff} = \alpha\left(1-\kappa_{\rm coll}\right) \,.
\end{eqnarray}
The parameter $\tau_{\rm sw}$ stands for the time when the sound wave period ends, after which the energy fluid motion is transferred to turbulence, with the expression \cite{Ellis:2020awk}
\begin{eqnarray}
\tau_{\rm sw}=\frac{R_*}{U_f}\,,\quad U_f=\sqrt{\frac{3}{4}\frac{\alpha_{\rm eff}}{1+\alpha_{\rm eff}}\kappa_{\rm sw}}\,,
\end{eqnarray}
where $U_f$ is the root-mean-square fluid velocity.

The spectral form functions $S_{\rm coll}$, $S_{\rm sw}$ and $S_{\rm turb}$ read \cite{Ellis:2020nnr}
\begin{subequations}
\begin{eqnarray}
S_{\rm coll}(f) &=&  \left[1 + 0.05 \left(\frac{f}{f_{\rm coll}}\right)^{-1.61}\right] \left(\frac{f}{f_{\rm coll}}\right)^{2.54}\left[1 + 1.13 \left(\frac{f}{f_{\rm coll}}\right)^{2.08}\right]^{-2.30}\,,\\
S_{\rm sw}(f) &=& \left(\frac{f}{f_{\rm sw}}\right)^3\left(1+\frac{3}{4}\frac{f^2}{f_{\rm sw}^2}\right)^{-\frac72}\,,\\
S_{\rm turb}(f) &=& \left(\frac{f}{f_{\rm turb}}\right)^3\left(1+\frac{f}{f_{\rm turb}}\right)^{-\frac{11}{3}} \left(1+8\pi \frac{f}{\Hub_*}\right)^{-1}\,.
\end{eqnarray}
\end{subequations}
The peak frequencies $f_{\rm coll}$, $f_{\rm sw}$ and $f_{\rm turb}$ are given by
\begin{subequations}
\begin{eqnarray}
f_{\rm coll} &=& 16.5\, \mu \text{Hz}\, \frac{0.62}{1.8 + 0.1v_w + v_w^2} \left(\frac{\beta}{\Hub_*}\right) \left(\frac{T_*}{100\,\text{GeV}}\right) \left(\frac{g_*}{100}\right)^{\frac16}\,,\\
f_{\rm sw} &=& 19\, \mu \text{Hz}\, \frac{1}{v_w} \left(\frac{\beta}{\Hub_*}\right) \left(\frac{T_*}{100\,\text{GeV}}\right) \left(\frac{g_*}{100}\right)^{\frac16}\,,\\
f_{\rm turb} &=& 27\, \mu \text{Hz}\, \frac{1}{v_w} \left(\frac{\beta}{\Hub_*}\right) \left(\frac{T_*}{100\,\text{GeV}}\right) \left(\frac{g_*}{100}\right)^{\frac16}\,.
\end{eqnarray}
\end{subequations}

\acknowledgments
BF acknowledges the Chinese Scholarship Council (CSC) Grant No.\ 201809210011 under agreements [2018]3101 and [2019]536. SFK acknowledges the STFC Consolidated Grant ST/L000296/1 and the European Union's Horizon 2020 Research and Innovation programme under Marie Sklodowska-Curie grant agreement HIDDeN European ITN project (H2020-MSCA-ITN-2019//860881-HIDDeN). 

%%%%%%%%%%%%%%%%%%%%%%%%%%%%%%%%%%%%%%%%%%%%%%%%%%%%%%%%%%%%%%%%%%%%%
\bibliographystyle{JHEP}
\bibliography{Ref}

\providecommand{\href}[2]{#2}\begingroup\raggedright\begin{thebibliography}{10}

\bibitem{Pati:1973uk}
J.C.~Pati and A.~Salam, \emph{{Unified Lepton-Hadron Symmetry and a Gauge
  Theory of the Basic Interactions}},
  \href{https://doi.org/10.1103/PhysRevD.8.1240}{\emph{Phys. Rev. D} {\bfseries
  8} (1973) 1240}.

\bibitem{Pati:1974yy}
J.C.~Pati and A.~Salam, \emph{{Lepton Number as the Fourth Color}},
  \href{https://doi.org/10.1103/PhysRevD.10.275}{\emph{Phys. Rev. D} {\bfseries
  10} (1974) 275}.

\bibitem{Senjanovic:1982ex}
G.~Senjanovic and A.~Sokorac, \emph{{Light Leptoquarks in SO(10)}},
  \href{https://doi.org/10.1007/BF01574858}{\emph{Z. Phys. C} {\bfseries 20}
  (1983) 255}.

\bibitem{Buchmuller:1986iq}
W.~Buchmuller and D.~Wyler, \emph{{Constraints on SU(5) Type Leptoquarks}},
  \href{https://doi.org/10.1016/0370-2693(86)90771-9}{\emph{Phys. Lett. B}
  {\bfseries 177} (1986) 377}.

\bibitem{Frampton:1991ay}
P.H.~Frampton, \emph{{Light leptoquarks as possible signature of strong
  electroweak unification}},
  \href{https://doi.org/10.1142/S0217732392000525}{\emph{Mod. Phys. Lett. A}
  {\bfseries 7} (1992) 559}.

\bibitem{Gershtein:1999gp}
S.S.~Gershtein, A.A.~Likhoded and A.I.~Onishchenko, \emph{{TeV-scale
  leptoquarks from GUTs/string/M-theory unification}},
  \href{https://doi.org/10.1016/S0370-1573(99)00063-0}{\emph{Phys. Rept.}
  {\bfseries 320} (1999) 159}.

\bibitem{Fuentes-Martin:2020pww}
J.~Fuentes-Martin, G.~Isidori, J.~Pag\`es and B.A.~Stefanek, \emph{{Flavor
  non-universal Pati-Salam unification and neutrino masses}},
  \href{https://doi.org/10.1016/j.physletb.2021.136484}{\emph{Phys. Lett. B}
  {\bfseries 820} (2021) 136484}
  [\href{https://arxiv.org/abs/2012.10492}{{\ttfamily 2012.10492}}].

\bibitem{King:2021jeo}
S.F.~King, \emph{{Twin Pati-Salam theory of flavour with a TeV scale vector
  leptoquark}}, \href{https://doi.org/10.1007/JHEP11(2021)161}{\emph{JHEP}
  {\bfseries 11} (2021) 161}
  [\href{https://arxiv.org/abs/2106.03876}{{\ttfamily 2106.03876}}].

\bibitem{FernandezNavarro:2022gst}
M.~Fern\'andez~Navarro and S.F.~King, \emph{{$B$-anomalies in a twin Pati-Salam
  theory of flavour}},  \href{https://arxiv.org/abs/2209.00276}{{\ttfamily
  2209.00276}}.

\bibitem{LHCb:2019hip}
{\scshape LHCb} collaboration, \emph{{Search for lepton-universality violation
  in $B^+\to K^+\ell^+\ell^-$ decays}},
  \href{https://doi.org/10.1103/PhysRevLett.122.191801}{\emph{Phys. Rev. Lett.}
  {\bfseries 122} (2019) 191801}
  [\href{https://arxiv.org/abs/1903.09252}{{\ttfamily 1903.09252}}].

\bibitem{Belle:2019rba}
{\scshape Belle} collaboration, \emph{{Measurement of $\mathcal{R}(D)$ and
  $\mathcal{R}(D^*)$ with a semileptonic tagging method}},
  \href{https://doi.org/10.1103/PhysRevLett.124.161803}{\emph{Phys. Rev. Lett.}
  {\bfseries 124} (2020) 161803}
  [\href{https://arxiv.org/abs/1910.05864}{{\ttfamily 1910.05864}}].

\bibitem{LHCb:2021trn}
{\scshape LHCb} collaboration, \emph{{Test of lepton universality in
  beauty-quark decays}},
  \href{https://doi.org/10.1038/s41567-021-01478-8}{\emph{Nature Phys.}
  {\bfseries 18} (2022) 277}
  [\href{https://arxiv.org/abs/2103.11769}{{\ttfamily 2103.11769}}].

\bibitem{LHCb:2021lvy}
{\scshape LHCb} collaboration, \emph{{Tests of lepton universality using
  $B^0\to K^0_S \ell^+ \ell^-$ and $B^+\to K^{*+} \ell^+ \ell^-$ decays}},
  \href{https://doi.org/10.1103/PhysRevLett.128.191802}{\emph{Phys. Rev. Lett.}
  {\bfseries 128} (2022) 191802}
  [\href{https://arxiv.org/abs/2110.09501}{{\ttfamily 2110.09501}}].

\bibitem{Angelescu:2021lln}
A.~Angelescu, D.~Be\v{c}irevi\'c, D.A.~Faroughy, F.~Jaffredo and O.~Sumensari,
  \emph{{Single leptoquark solutions to the B-physics anomalies}},
  \href{https://doi.org/10.1103/PhysRevD.104.055017}{\emph{Phys. Rev. D}
  {\bfseries 104} (2021) 055017}
  [\href{https://arxiv.org/abs/2103.12504}{{\ttfamily 2103.12504}}].

\bibitem{Becirevic:2022tsj}
D.~Be\v{c}irevi\'c, I.~Dor\v{s}ner, S.~Fajfer, D.A.~Faroughy, F.~Jaffredo,
  N.~Ko\v{s}nik et~al., \emph{{On a model with two scalar leptoquarks --$R_2$
  and $S_3$}},  \href{https://arxiv.org/abs/2206.09717}{{\ttfamily
  2206.09717}}.

\bibitem{Cheung:2001ip}
K.-m.~Cheung, \emph{{Muon anomalous magnetic moment and leptoquark solutions}},
  \href{https://doi.org/10.1103/PhysRevD.64.033001}{\emph{Phys. Rev. D}
  {\bfseries 64} (2001) 033001}
  [\href{https://arxiv.org/abs/hep-ph/0102238}{{\ttfamily hep-ph/0102238}}].

\bibitem{ColuccioLeskow:2016dox}
E.~Coluccio~Leskow, G.~D'Ambrosio, A.~Crivellin and D.~M\"uller,
  \emph{{$(g-2)\mu$, lepton flavor violation, and $Z$ decays with leptoquarks:
  Correlations and future prospects}},
  \href{https://doi.org/10.1103/PhysRevD.95.055018}{\emph{Phys. Rev. D}
  {\bfseries 95} (2017) 055018}
  [\href{https://arxiv.org/abs/1612.06858}{{\ttfamily 1612.06858}}].

\bibitem{Crivellin:2019dwb}
A.~Crivellin, D.~M\"uller and F.~Saturnino, \emph{{Flavor Phenomenology of the
  Leptoquark Singlet-Triplet Model}},
  \href{https://doi.org/10.1007/JHEP06(2020)020}{\emph{JHEP} {\bfseries 06}
  (2020) 020} [\href{https://arxiv.org/abs/1912.04224}{{\ttfamily
  1912.04224}}].

\bibitem{Athron:2021iuf}
P.~Athron, C.~Bal\'azs, D.H.J.~Jacob, W.~Kotlarski, D.~St\"ockinger and
  H.~St\"ockinger-Kim, \emph{{New physics explanations of $a_\mu$ in light of
  the FNAL muon g \ensuremath{-} 2 measurement}},
  \href{https://doi.org/10.1007/JHEP09(2021)080}{\emph{JHEP} {\bfseries 09}
  (2021) 080} [\href{https://arxiv.org/abs/2104.03691}{{\ttfamily
  2104.03691}}].

\bibitem{Du:2021zkq}
M.~Du, J.~Liang, Z.~Liu and V.Q.~Tran, \emph{{A vector leptoquark
  interpretation of the muon $g-2$ and $B$ anomalies}},
  \href{https://arxiv.org/abs/2104.05685}{{\ttfamily 2104.05685}}.

\bibitem{Chen:2022hle}
S.-L.~Chen, W.-w.~Jiang and Z.-K.~Liu, \emph{{Combined explanations of
  $B$-physics anomalies, $(g-2)_{e, \mu}$ and neutrino masses by scalar
  leptoquarks}},  \href{https://arxiv.org/abs/2205.15794}{{\ttfamily
  2205.15794}}.

\bibitem{Mahanta:1999xd}
U.~Mahanta, \emph{{Neutrino masses and mixing angles from leptoquark
  interactions}}, \href{https://doi.org/10.1103/PhysRevD.62.073009}{\emph{Phys.
  Rev. D} {\bfseries 62} (2000) 073009}
  [\href{https://arxiv.org/abs/hep-ph/9909518}{{\ttfamily hep-ph/9909518}}].

\bibitem{Deppisch:2016qqd}
F.F.~Deppisch, S.~Kulkarni, H.~P\"as and E.~Schumacher, \emph{{Leptoquark
  patterns unifying neutrino masses, flavor anomalies, and the diphoton
  excess}}, \href{https://doi.org/10.1103/PhysRevD.94.013003}{\emph{Phys. Rev.
  D} {\bfseries 94} (2016) 013003}
  [\href{https://arxiv.org/abs/1603.07672}{{\ttfamily 1603.07672}}].

\bibitem{Popov:2016fzr}
O.~Popov and G.A.~White, \emph{{One Leptoquark to unify them? Neutrino masses
  and unification in the light of $(g-2)_\mu$, $R_{D^{(\star)}}$ and $R_K$
  anomalies}},
  \href{https://doi.org/10.1016/j.nuclphysb.2017.08.007}{\emph{Nucl. Phys. B}
  {\bfseries 923} (2017) 324}
  [\href{https://arxiv.org/abs/1611.04566}{{\ttfamily 1611.04566}}].

\bibitem{Cai:2017wry}
Y.~Cai, J.~Gargalionis, M.A.~Schmidt and R.R.~Volkas, \emph{{Reconsidering the
  One Leptoquark solution: flavor anomalies and neutrino mass}},
  \href{https://doi.org/10.1007/JHEP10(2017)047}{\emph{JHEP} {\bfseries 10}
  (2017) 047} [\href{https://arxiv.org/abs/1704.05849}{{\ttfamily
  1704.05849}}].

\bibitem{BhupalDev:2020zcy}
P.S.~Bhupal~Dev, R.~Mohanta, S.~Patra and S.~Sahoo, \emph{{Unified explanation
  of flavor anomalies, radiative neutrino masses, and ANITA anomalous events in
  a vector leptoquark model}},
  \href{https://doi.org/10.1103/PhysRevD.102.095012}{\emph{Phys. Rev. D}
  {\bfseries 102} (2020) 095012}
  [\href{https://arxiv.org/abs/2004.09464}{{\ttfamily 2004.09464}}].

\bibitem{Crivellin:2020ukd}
A.~Crivellin, D.~M\"uller and F.~Saturnino, \emph{{Leptoquarks in oblique
  corrections and Higgs signal strength: status and prospects}},
  \href{https://doi.org/10.1007/JHEP11(2020)094}{\emph{JHEP} {\bfseries 11}
  (2020) 094} [\href{https://arxiv.org/abs/2006.10758}{{\ttfamily
  2006.10758}}].

\bibitem{DAlise:2022ypp}
A.~D'Alise et~al., \emph{{Standard model anomalies: lepton flavour
  non-universality, g \ensuremath{-} 2 and W-mass}},
  \href{https://doi.org/10.1007/JHEP08(2022)125}{\emph{JHEP} {\bfseries 08}
  (2022) 125} [\href{https://arxiv.org/abs/2204.03686}{{\ttfamily
  2204.03686}}].

\bibitem{Athron:2022qpo}
P.~Athron, A.~Fowlie, C.-T.~Lu, L.~Wu, Y.~Wu and B.~Zhu, \emph{{The $W$ boson
  Mass and Muon $g-2$: Hadronic Uncertainties or New Physics?}},
  \href{https://arxiv.org/abs/2204.03996}{{\ttfamily 2204.03996}}.

\bibitem{Cheung:2022zsb}
K.~Cheung, W.-Y.~Keung and P.-Y.~Tseng, \emph{{Isodoublet vector leptoquark
  solution to the muon g-2, RK,K*, RD,D*, and W-mass anomalies}},
  \href{https://doi.org/10.1103/PhysRevD.106.015029}{\emph{Phys. Rev. D}
  {\bfseries 106} (2022) 015029}
  [\href{https://arxiv.org/abs/2204.05942}{{\ttfamily 2204.05942}}].

\bibitem{Bhaskar:2022vgk}
A.~Bhaskar, A.A.~Madathil, T.~Mandal and S.~Mitra, \emph{{Combined explanation
  of $W$-mass, muon $g-2$, $R_{K^{(*)}}$ and $R_{D^{(*)}}$ anomalies in a
  singlet-triplet scalar leptoquark model}},
  \href{https://arxiv.org/abs/2204.09031}{{\ttfamily 2204.09031}}.

\bibitem{He:2022zjz}
S.-P.~He, \emph{{A leptoquark and vector-like quark extended model for the
  simultaneous explanation of the $W$ boson mass and muon $g-2$ anomalies}},
  \href{https://arxiv.org/abs/2205.02088}{{\ttfamily 2205.02088}}.

\bibitem{Kolb:1997rb}
S.~Kolb, M.~Hirsch and H.V.~Klapdor-Kleingrothaus, \emph{{Bounds on leptoquark
  parameters with nonvanishing leptoquark Higgs couplings}},
  \href{https://doi.org/10.1016/S0370-2693(96)01445-1}{\emph{Phys. Lett. B}
  {\bfseries 391} (1997) 131}.

\bibitem{Dorsner:2016wpm}
I.~Dor\v{s}ner, S.~Fajfer, A.~Greljo, J.F.~Kamenik and N.~Ko\v{s}nik,
  \emph{{Physics of leptoquarks in precision experiments and at particle
  colliders}}, \href{https://doi.org/10.1016/j.physrep.2016.06.001}{\emph{Phys.
  Rept.} {\bfseries 641} (2016) 1}
  [\href{https://arxiv.org/abs/1603.04993}{{\ttfamily 1603.04993}}].

\bibitem{Curtin:2014jma}
D.~Curtin, P.~Meade and C.-T.~Yu, \emph{{Testing Electroweak Baryogenesis with
  Future Colliders}},
  \href{https://doi.org/10.1007/JHEP11(2014)127}{\emph{JHEP} {\bfseries 11}
  (2014) 127} [\href{https://arxiv.org/abs/1409.0005}{{\ttfamily 1409.0005}}].

\bibitem{Caprini:2015zlo}
C.~Caprini et~al., \emph{{Science with the space-based interferometer eLISA.
  II: Gravitational waves from cosmological phase transitions}},
  \href{https://doi.org/10.1088/1475-7516/2016/04/001}{\emph{JCAP} {\bfseries
  04} (2016) 001} [\href{https://arxiv.org/abs/1512.06239}{{\ttfamily
  1512.06239}}].

\bibitem{Weir:2017wfa}
D.J.~Weir, \emph{{Gravitational waves from a first order electroweak phase
  transition: a brief review}},
  \href{https://doi.org/10.1098/rsta.2017.0126}{\emph{Phil. Trans. Roy. Soc.
  Lond. A} {\bfseries 376} (2018) 20170126}
  [\href{https://arxiv.org/abs/1705.01783}{{\ttfamily 1705.01783}}].

\bibitem{Caprini:2019egz}
C.~Caprini et~al., \emph{{Detecting gravitational waves from cosmological phase
  transitions with LISA: an update}},
  \href{https://doi.org/10.1088/1475-7516/2020/03/024}{\emph{JCAP} {\bfseries
  03} (2020) 024} [\href{https://arxiv.org/abs/1910.13125}{{\ttfamily
  1910.13125}}].

\bibitem{Beniwal:2017eik}
A.~Beniwal, M.~Lewicki, J.D.~Wells, M.~White and A.G.~Williams,
  \emph{{Gravitational wave, collider and dark matter signals from a scalar
  singlet electroweak baryogenesis}},
  \href{https://doi.org/10.1007/JHEP08(2017)108}{\emph{JHEP} {\bfseries 08}
  (2017) 108} [\href{https://arxiv.org/abs/1702.06124}{{\ttfamily
  1702.06124}}].

\bibitem{DiBari:2021dri}
P.~Di~Bari, D.~Marfatia and Y.-L.~Zhou, \emph{{Gravitational waves from
  first-order phase transitions in Majoron models of neutrino mass}},
  \href{https://doi.org/10.1007/JHEP10(2021)193}{\emph{JHEP} {\bfseries 10}
  (2021) 193} [\href{https://arxiv.org/abs/2106.00025}{{\ttfamily
  2106.00025}}].

\bibitem{Bian:2021dmp}
L.~Bian, Y.-L.~Tang and R.~Zhou, \emph{{FIMP dark matter mediated by a massive
  gauge boson around the phase transition period and the gravitational waves
  production}}, \href{https://doi.org/10.1103/PhysRevD.106.035028}{\emph{Phys.
  Rev. D} {\bfseries 106} (2022) 035028}
  [\href{https://arxiv.org/abs/2111.10608}{{\ttfamily 2111.10608}}].

\bibitem{Demidov:2021lyo}
S.~Demidov, D.~Gorbunov and E.~Kriukova, \emph{{Gravitational waves from
  first-order electroweak phase transition in a model with light sgoldstinos}},
  \href{https://doi.org/10.1007/JHEP07(2022)061}{\emph{JHEP} {\bfseries 07}
  (2022) 061} [\href{https://arxiv.org/abs/2112.06083}{{\ttfamily
  2112.06083}}].

\bibitem{Graf:2021xku}
L.~Gr\'af, S.~Jana, A.~Kaladharan and S.~Saad, \emph{{Gravitational wave
  imprints of left-right symmetric model with minimal Higgs sector}},
  \href{https://doi.org/10.1088/1475-7516/2022/05/003}{\emph{JCAP} {\bfseries
  05} (2022) 003} [\href{https://arxiv.org/abs/2112.12041}{{\ttfamily
  2112.12041}}].

\bibitem{Zhou:2022mlz}
R.~Zhou, L.~Bian and Y.~Du, \emph{{Electroweak phase transition and
  gravitational waves in the type-II seesaw model}},
  \href{https://doi.org/10.1007/JHEP08(2022)205}{\emph{JHEP} {\bfseries 08}
  (2022) 205} [\href{https://arxiv.org/abs/2203.01561}{{\ttfamily
  2203.01561}}].

\bibitem{Chen:2022zsh}
T.-K.~Chen, C.-W.~Chiang, C.-T.~Huang and B.-Q.~Lu, \emph{{Updated constraints
  on the Georgi-Machacek model and its electroweak phase transition and
  associated gravitational waves}},
  \href{https://doi.org/10.1103/PhysRevD.106.055019}{\emph{Phys. Rev. D}
  {\bfseries 106} (2022) 055019}
  [\href{https://arxiv.org/abs/2205.02064}{{\ttfamily 2205.02064}}].

\bibitem{Lu:2022zpn}
B.-Q.~Lu, C.-W.~Chiang and D.~Huang, \emph{{Probing WIMPs in space-based
  gravitational wave experiments}},
  \href{https://doi.org/10.1016/j.physletb.2022.137308}{\emph{Phys. Lett. B}
  {\bfseries 833} (2022) 137308}
  [\href{https://arxiv.org/abs/2205.08380}{{\ttfamily 2205.08380}}].

\bibitem{Dasgupta:2022isg}
A.~Dasgupta, P.S.B.~Dev, A.~Ghoshal and A.~Mazumdar, \emph{{Gravitational wave
  pathway to testable leptogenesis}},
  \href{https://doi.org/10.1103/PhysRevD.106.075027}{\emph{Phys. Rev. D}
  {\bfseries 106} (2022) 075027}
  [\href{https://arxiv.org/abs/2206.07032}{{\ttfamily 2206.07032}}].

\bibitem{Niemi:2021qvp}
L.~Niemi, P.~Schicho and T.V.I.~Tenkanen, \emph{{Singlet-assisted electroweak
  phase transition at two loops}},
  \href{https://doi.org/10.1103/PhysRevD.103.115035}{\emph{Phys. Rev. D}
  {\bfseries 103} (2021) 115035}
  [\href{https://arxiv.org/abs/2103.07467}{{\ttfamily 2103.07467}}].

\bibitem{Schicho:2022wty}
P.~Schicho, T.V.I.~Tenkanen and G.~White, \emph{{Combining thermal resummation
  and gauge invariance for electroweak phase transition}},
  \href{https://arxiv.org/abs/2203.04284}{{\ttfamily 2203.04284}}.

\bibitem{Quiros:1999jp}
M.~Quiros, \emph{{Finite temperature field theory and phase transitions}},  in
  \emph{{ICTP Summer School in High-Energy Physics and Cosmology}},
  pp.~187--259, 1, 1999 [\href{https://arxiv.org/abs/hep-ph/9901312}{{\ttfamily
  hep-ph/9901312}}].

\bibitem{Linde:1980ts}
A.D.~Linde, \emph{{Infrared Problem in Thermodynamics of the Yang-Mills Gas}},
  \href{https://doi.org/10.1016/0370-2693(80)90769-8}{\emph{Phys. Lett. B}
  {\bfseries 96} (1980) 289}.

\bibitem{Croon:2020cgk}
D.~Croon, O.~Gould, P.~Schicho, T.V.I.~Tenkanen and G.~White,
  \emph{{Theoretical uncertainties for cosmological first-order phase
  transitions}}, \href{https://doi.org/10.1007/JHEP04(2021)055}{\emph{JHEP}
  {\bfseries 04} (2021) 055}
  [\href{https://arxiv.org/abs/2009.10080}{{\ttfamily 2009.10080}}].

\bibitem{Parwani:1991gq}
R.R.~Parwani, \emph{{Resummation in a hot scalar field theory}},
  \href{https://doi.org/10.1103/PhysRevD.45.4695}{\emph{Phys. Rev. D}
  {\bfseries 45} (1992) 4695}
  [\href{https://arxiv.org/abs/hep-ph/9204216}{{\ttfamily hep-ph/9204216}}].

\bibitem{Curtin:2016urg}
D.~Curtin, P.~Meade and H.~Ramani, \emph{{Thermal Resummation and Phase
  Transitions}},
  \href{https://doi.org/10.1140/epjc/s10052-018-6268-0}{\emph{Eur. Phys. J. C}
  {\bfseries 78} (2018) 787}
  [\href{https://arxiv.org/abs/1612.00466}{{\ttfamily 1612.00466}}].

\bibitem{Arnold:1992rz}
P.B.~Arnold and O.~Espinosa, \emph{{The Effective potential and first order
  phase transitions: Beyond leading-order}},
  \href{https://doi.org/10.1103/PhysRevD.47.3546}{\emph{Phys. Rev. D}
  {\bfseries 47} (1993) 3546}
  [\href{https://arxiv.org/abs/hep-ph/9212235}{{\ttfamily hep-ph/9212235}}].

\bibitem{Cline:2011mm}
J.M.~Cline, K.~Kainulainen and M.~Trott, \emph{{Electroweak Baryogenesis in Two
  Higgs Doublet Models and B meson anomalies}},
  \href{https://doi.org/10.1007/JHEP11(2011)089}{\emph{JHEP} {\bfseries 11}
  (2011) 089} [\href{https://arxiv.org/abs/1107.3559}{{\ttfamily 1107.3559}}].

\bibitem{Hiller:2017bzc}
G.~Hiller and I.~Nisandzic, \emph{{$R_K$ and $R_{K^{\ast}}$ beyond the standard
  model}}, \href{https://doi.org/10.1103/PhysRevD.96.035003}{\emph{Phys. Rev.
  D} {\bfseries 96} (2017) 035003}
  [\href{https://arxiv.org/abs/1704.05444}{{\ttfamily 1704.05444}}].

\bibitem{Weinberg:1987vp}
E.J.~Weinberg and A.-q.~Wu, \emph{{UNDERSTANDING COMPLEX PERTURBATIVE EFFECTIVE
  POTENTIALS}}, \href{https://doi.org/10.1103/PhysRevD.36.2474}{\emph{Phys.
  Rev. D} {\bfseries 36} (1987) 2474}.

\bibitem{TheATLASandCMSCollaborations:2015bln}
\emph{{Measurements of the Higgs boson production and decay rates and
  constraints on its couplings from a combined ATLAS and CMS analysis of the
  LHC pp collision data at $\sqrt{s}$ = 7 and 8 TeV}}, .

\bibitem{CMS:2015kwa}
{\scshape CMS} collaboration, \emph{{Measurements of the Higgs boson production
  and decay rates and constraints on its couplings from a combined ATLAS and
  CMS analysis of the LHC pp collision data at sqrt s = 7 and 8 TeV}}, .

\bibitem{ATLAS:2022tnm}
{\scshape ATLAS} collaboration, \emph{{Measurement of the properties of Higgs
  boson production at $\sqrt{s} = 13$ TeV in the $H\to\gamma\gamma$ channel
  using $139$ fb$^{-1}$ of $pp$ collision data with the ATLAS experiment}},
  \href{https://arxiv.org/abs/2207.00348}{{\ttfamily 2207.00348}}.

\bibitem{CMS:2022dwd}
{\scshape CMS} collaboration, \emph{{A portrait of the Higgs boson by the CMS
  experiment ten years after the discovery}},
  \href{https://doi.org/10.1038/s41586-022-04892-x}{\emph{Nature} {\bfseries
  607} (2022) 60} [\href{https://arxiv.org/abs/2207.00043}{{\ttfamily
  2207.00043}}].

\bibitem{ParticleDataGroup:2020ssz}
{\scshape Particle Data Group} collaboration, \emph{{Review of Particle
  Physics}}, \href{https://doi.org/10.1093/ptep/ptaa104}{\emph{PTEP} {\bfseries
  2020} (2020) 083C01}.

\bibitem{CMS:2018qgz}
{\scshape CMS} collaboration, \emph{{Sensitivity projections for Higgs boson
  properties measurements at the HL-LHC}}, .

\bibitem{dEnterria:2017dac}
D.~d'Enterria, \emph{{Higgs physics at the Future Circular Collider}},
  \href{https://doi.org/10.22323/1.282.0434}{\emph{PoS} {\bfseries ICHEP2016}
  (2017) 434} [\href{https://arxiv.org/abs/1701.02663}{{\ttfamily
  1701.02663}}].

\bibitem{FCC:2018evy}
{\scshape FCC} collaboration, \emph{{FCC-ee: The Lepton Collider}: {Future
  Circular Collider Conceptual Design Report Volume 2}},
  \href{https://doi.org/10.1140/epjst/e2019-900045-4}{\emph{Eur. Phys. J. ST}
  {\bfseries 228} (2019) 261}.

\bibitem{Bambade:2019fyw}
P.~Bambade et~al., \emph{{The International Linear Collider: A Global
  Project}},  \href{https://arxiv.org/abs/1903.01629}{{\ttfamily 1903.01629}}.

\bibitem{An:2018dwb}
F.~An et~al., \emph{{Precision Higgs physics at the CEPC}},
  \href{https://doi.org/10.1088/1674-1137/43/4/043002}{\emph{Chin. Phys. C}
  {\bfseries 43} (2019) 043002}
  [\href{https://arxiv.org/abs/1810.09037}{{\ttfamily 1810.09037}}].

\bibitem{Ruan:2021gap}
M.~Ruan, Y.~Fang, G.~Li and D.~Yu, \emph{{CEPC Research Report: Higgs Physics
  Analysis}},  \href{https://arxiv.org/abs/2107.09820}{{\ttfamily 2107.09820}}.

\bibitem{Crivellin:2020mjs}
A.~Crivellin, C.~Greub, D.~M\"uller and F.~Saturnino, \emph{{Scalar Leptoquarks
  in Leptonic Processes}},
  \href{https://doi.org/10.1007/JHEP02(2021)182}{\emph{JHEP} {\bfseries 02}
  (2021) 182} [\href{https://arxiv.org/abs/2010.06593}{{\ttfamily
  2010.06593}}].

\bibitem{Dine:1992wr}
M.~Dine, R.G.~Leigh, P.Y.~Huet, A.D.~Linde and D.A.~Linde, \emph{{Towards the
  theory of the electroweak phase transition}},
  \href{https://doi.org/10.1103/PhysRevD.46.550}{\emph{Phys. Rev. D} {\bfseries
  46} (1992) 550} [\href{https://arxiv.org/abs/hep-ph/9203203}{{\ttfamily
  hep-ph/9203203}}].

\bibitem{CMS:2020gru}
{\scshape CMS} collaboration, \emph{{Search for singly and pair-produced
  leptoquarks coupling to third-generation fermions in proton-proton collisions
  at $\sqrt{s}$ = 13 TeV}}, .

\bibitem{CMS:2020wzx}
{\scshape CMS} collaboration, \emph{{Search for singly and pair-produced
  leptoquarks coupling to third-generation fermions in proton-proton collisions
  at s=13~TeV}},
  \href{https://doi.org/10.1016/j.physletb.2021.136446}{\emph{Phys. Lett. B}
  {\bfseries 819} (2021) 136446}
  [\href{https://arxiv.org/abs/2012.04178}{{\ttfamily 2012.04178}}].

\bibitem{ATLAS:2021oiz}
{\scshape ATLAS} collaboration, \emph{{Search for pair production of
  third-generation scalar leptoquarks decaying into a top quark and a
  $\tau$-lepton in $pp$ collisions at $ \sqrt{s} $ = 13 TeV with the ATLAS
  detector}}, \href{https://doi.org/10.1007/JHEP06(2021)179}{\emph{JHEP}
  {\bfseries 06} (2021) 179}
  [\href{https://arxiv.org/abs/2101.11582}{{\ttfamily 2101.11582}}].

\bibitem{ATLAS:2022fho}
{\scshape ATLAS} collaboration, \emph{{Search for scalar leptoquarks in the
  b$\tau\tau$ final state in $pp$ collisions at $\sqrt{s}
  =$\textasciitilde{}13\textasciitilde{}TeV with the ATLAS detector}}, .

\bibitem{Corbin:2005ny}
V.~Corbin and N.J.~Cornish, \emph{{Detecting the cosmic gravitational wave
  background with the big bang observer}},
  \href{https://doi.org/10.1088/0264-9381/23/7/014}{\emph{Class. Quant. Grav.}
  {\bfseries 23} (2006) 2435}
  [\href{https://arxiv.org/abs/gr-qc/0512039}{{\ttfamily gr-qc/0512039}}].

\bibitem{Kawamura:2019jqt}
{\scshape DECIGO working group} collaboration, \emph{{Primordial gravitational
  wave and DECIGO}}, \href{https://doi.org/10.22323/1.356.0019}{\emph{PoS}
  {\bfseries KMI2019} (2019) 019}.

\bibitem{Kawamura:2020pcg}
S.~Kawamura et~al., \emph{{Current status of space gravitational wave antenna
  DECIGO and B-DECIGO}},
  \href{https://doi.org/10.1093/ptep/ptab019}{\emph{PTEP} {\bfseries 2021}
  (2021) 05A105} [\href{https://arxiv.org/abs/2006.13545}{{\ttfamily
  2006.13545}}].

\bibitem{Sesana:2019vho}
A.~Sesana et~al., \emph{{Unveiling the gravitational universe at $\mu$-Hz
  frequencies}}, \href{https://doi.org/10.1007/s10686-021-09709-9}{\emph{Exper.
  Astron.} {\bfseries 51} (2021) 1333}
  [\href{https://arxiv.org/abs/1908.11391}{{\ttfamily 1908.11391}}].

\bibitem{Ellis:2018mja}
J.~Ellis, M.~Lewicki and J.M.~No, \emph{{On the Maximal Strength of a
  First-Order Electroweak Phase Transition and its Gravitational Wave Signal}},
  \href{https://doi.org/10.1088/1475-7516/2019/04/003}{\emph{JCAP} {\bfseries
  04} (2019) 003} [\href{https://arxiv.org/abs/1809.08242}{{\ttfamily
  1809.08242}}].

\bibitem{Ellis:2020nnr}
J.~Ellis, M.~Lewicki and V.~Vaskonen, \emph{{Updated predictions for
  gravitational waves produced in a strongly supercooled phase transition}},
  \href{https://doi.org/10.1088/1475-7516/2020/11/020}{\emph{JCAP} {\bfseries
  11} (2020) 020} [\href{https://arxiv.org/abs/2007.15586}{{\ttfamily
  2007.15586}}].

\bibitem{Ellis:2022lft}
J.~Ellis, M.~Lewicki, M.~Merchand, J.M.~No and M.~Zych, \emph{{The scalar
  singlet extension of the Standard Model: gravitational waves versus
  baryogenesis}}, \href{https://doi.org/10.1007/JHEP01(2023)093}{\emph{JHEP}
  {\bfseries 01} (2023) 093}
  [\href{https://arxiv.org/abs/2210.16305}{{\ttfamily 2210.16305}}].

\bibitem{Lewicki:2021pgr}
M.~Lewicki, M.~Merchand and M.~Zych, \emph{{Electroweak bubble wall expansion:
  gravitational waves and baryogenesis in Standard Model-like thermal plasma}},
  \href{https://doi.org/10.1007/JHEP02(2022)017}{\emph{JHEP} {\bfseries 02}
  (2022) 017} [\href{https://arxiv.org/abs/2111.02393}{{\ttfamily
  2111.02393}}].

\bibitem{Ellis:2019oqb}
J.~Ellis, M.~Lewicki, J.M.~No and V.~Vaskonen, \emph{{Gravitational wave energy
  budget in strongly supercooled phase transitions}},
  \href{https://doi.org/10.1088/1475-7516/2019/06/024}{\emph{JCAP} {\bfseries
  06} (2019) 024} [\href{https://arxiv.org/abs/1903.09642}{{\ttfamily
  1903.09642}}].

\bibitem{Ellis:2020awk}
J.~Ellis, M.~Lewicki and J.M.~No, \emph{{Gravitational waves from first-order
  cosmological phase transitions: lifetime of the sound wave source}},
  \href{https://doi.org/10.1088/1475-7516/2020/07/050}{\emph{JCAP} {\bfseries
  07} (2020) 050} [\href{https://arxiv.org/abs/2003.07360}{{\ttfamily
  2003.07360}}].

\end{thebibliography}\endgroup
%%%%%%%%%%%%%%%%%%%%%%%%%%%%%%%%%%%%%%%%%%%%%%%%%%%%%%%%%%%%%%%%%%%%%

\end{document}